\documentclass[12pt]{article}
\usepackage{graphicx}
\usepackage{hyperref}
\usepackage{newtxtext,newtxmath}

\usepackage{scicite}

\usepackage{times}

\newcommand{\St}{\mathrm{St}}
\renewcommand{\v}[1]{{\boldsymbol{#1}}} 
\newcommand{\del}{\boldsymbol{\nabla}}
\newcommand{\grad}{\del}
\newcommand{\Div}{\del\cdot}
\newcommand{\cv}{c_{_{V}}}
\newcommand{\ptderiv}[1]{\frac{\partial{#1}}{\partial{t}}}

\usepackage[usenames,dvipsnames]{xcolor}
\usepackage{soul}

\newenvironment{sciabstract}{%
\begin{quote} \bf}
{\end{quote}}

\newcounter{lastnote}

\title{A planetesimal orbiting within the debris disc \\ around a white dwarf star}

\author{
Christopher J. Manser$^1$, 
Boris T. G\"ansicke$^{1,2}$,
Siegfried Eggl$^{3}$,
Mark Hollands$^1$,
Paula Izquierdo$^{4,5}$,\\
Detlev Koester$^6$,
John D. Landstreet$^{7,8}$,
Wladimir Lyra$^{3,9}$, 
Thomas R. Marsh$^1$,
Farzana Meru$^1$,\\
Alexander J. Mustill$^{10}$, 
Pablo Rodr\'iguez-Gil$^{4,5}$, 
Odette Toloza$^1$,
Dimitri Veras$^{1,2}$,\\
David J. Wilson$^{1,11}$,
Matthew R. Burleigh$^{12}$, 
Melvyn B. Davies$^{10}$,
Jay Farihi$^{13}$,
Nicola Gentile Fusillo$^{1}$,\\
Domitilla de Martino$^{14}$, 
Steven G. Parsons$^{15}$,
Andreas Quirrenbach$^{16}$, 
Roberto Raddi$^{17}$,\\
Sabine Reffert$^{16}$, 
Melania Del Santo$^{18}$,
Matthias R. Schreiber$^{19, 20}$, 
Roberto Silvotti$^{21}$, \\
Silvia Toonen$^{22, *}$, 
Eva Villaver$^{23}$, 
Mark Wyatt$^{24}$, 
Siyi Xu$^{25}$,
Simon Portegies Zwart$^{26}$\\
\small{$^{1}$ Department of Physics, University of Warwick, Coventry CV4 7AL, UK}\\
\small{$^{2}$ Centre for Exoplanets and Habitability, University of Warwick, Coventry CV4 7AL, UK}\\
\small{$^{3}$ Jet Propulsion Laboratory, California Institute of Technology, 4800 Oak Grove Drive, 91109 Pasadena, CA, USA}\\
\small{$^{4}$ Instituto de Astrof\'isica de Canarias, E-38205 La Laguna, Tenerife, Spain}\\
\small{$^{5}$ Universidad de La Laguna, Departamento de Astrof\'isica, E-38206 La Laguna, Tenerife, Spain}\\
\small{$^{6}$ Institut f\"{u}r Theoretische Physik und Astrophysik, Universit\"{a}t Kiel, 24098 Kiel, Germany}\\
\small{$^{7}$ Department of Physics and Astronomy, The University of Western Ontario, London, Ontario, N6A 3K7, Canada}\\
\small{$^{8}$} Armagh Observatory and Planetarium, College Hill, Armagh, Co. Armagh, BT61 9DG, UK\\
\small{$^{9}$ California State University, Northridge, Department of Physics and Astronomy, 18111 Nordhoff St, Northridge, CA, 91330}\\
\small{$^{10}$ Lund Observatory, Department of Astronomy \& Theoretical Physics, Lund University, Box 43, SE-221 00 Lund, Sweden}\\
\small{$^{11}$ McDonald Observatory, University of Texas at Austin, Austin, TX 78712}\\
\small{$^{12}$ Dept.\ of Physics and Astronomy, Leicester Institute of Space and Earth Observation,} \\ \small{University of Leicester, University Road, Leicester, LE1 7RH, UK} \\
\small{$^{13}$ Physics and Astronomy, University College London, London, WC1E 6BT, UK}\\
\small{$^{14}$ National Institute for Astrophysics, Osservatorio Astronomico di Capodimonte, Via Moiarello 16, 80131 Napoli, Italy}\\
\small{$^{15}$ The University of Sheffield, Western Bank, Sheffield, S10 2TN, UK}\\
\small{$^{16}$ Landessternwarte, Zentrum f\"ur Astronomie der Universit\"at Heidelberg, K\"{o}nigstuhl 12, 69117 Heidelberg, Germany}\\
\small{$^{17}$ Dr. Karl Remeis-Sternwarte, Astronomisches Institut der Universit\"at Erlangen-N\"urnberg, Sternwartestr. 7, 96049, Bamberg}\\
\small{$^{18}$ National Institute for Astrophysics/Institute of Space Astrophysics and Cosmic Physics, via Ugo La Malfa 153, 90146, Palermo, Italy}\\
\small{$^{19}$ Instituto de F\'isica y Astronom\'ia, Universidad de Valpara\'iso, Av. Gran Breta$\tilde{\textrm{n}}$a 1111, 5030 Casilla, Valpara\'iso, Chile}\\
\small{$^{20}$ Milennium Nucleus for Planet Formation - NPF, Universidad de Valpara\'iso, Av. Gran Breta$\tilde{\textrm{n}}$a 1111, Valpara\'iso, Chile}\\
\small{$^{21}$ National Institute for Astrophysics, Osservatorio Astrofisico di Torino, Strada dell'Osservatorio 20, 10025 Pino Torinese, Italy}\\
\small{$^{22}$ Anton Pannekoek Instituut voor Sterrenkunde, University of Amsterdam, P.O.Box 94249, 1090 GE, Amsterdam, The Netherlands}\\
\small{$^{23}$ Departamento de F\'isica Te\'orica, Universidad Aut\'onoma de Madrid, Cantoblanco 28049 Madrid, Spain}\\
\small{$^{24}$ Institute of Astronomy, Madingley Rd, Cambridge CB3 0HA, UK}\\
\small{$^{25}$ Gemini Observatory, Northern Operations Center, 670 N. A'ohoku Place, Hilo, Hawaii, 96720, USA}\\
\small{$^{26}$ Sterrewacht Leiden, Leiden University, P.O. Box 9513, 2300 RA Leiden, The Netherlands}\\
\small{$^{*}$ Present address: Institute for Gravitational Wave Astronomy and School of}\\ \small{Physics and Astronomy, University of Birmingham, Birmingham, B15 2TT, United Kingdom.}\\
}

\date{}

\begin{document} 

\baselineskip24pt

\maketitle 

\newpage
\hoffset0cm



\begin{sciabstract}
Many white dwarf stars show signs of having accreted smaller bodies, implying that they may host planetary systems. A small number of these systems contain gaseous debris discs, visible through emission lines. We report a stable 123.4\,min periodic variation in the strength and shape of the Ca\,\textsc{ii} emission line profiles originating from the debris disc around the white dwarf SDSS\,J122859.93+104032.9. We interpret this short-period signal as the signature of a solid body held together by its internal strength.
\end{sciabstract}


\noindent
Over 3000 planet-hosting stars are known \cite{hanetal14-1}, the vast majority of which will end their lives as white dwarfs. Theoretical models indicate that planetary systems, including the Solar System, can survive the evolution of their host star largely intact \cite{duncan+lissauer98-1, villaver+livio07-1, veras+gaensicke15-1}. Remnants of planetary systems have been indirectly detected in white dwarf systems via (i) the contaminated atmospheres of 25-50\% of white dwarfs, arising from the accretion of planetary material \cite{zuckermanetal10-1, koesteretal14-1}, (ii) compact dust discs \cite{zuckerman+becklin87-1,farihietal09-1}, formed from the rubble of tidally disrupted planetesimals \cite{jura03-1, verasetal14-1}, and (iii) atomic emission lines from gaseous discs co-located with the circumstellar dust \cite{gaensickeetal06-3, guoetal15-1}. The most direct evidence for remnant planetary systems around white dwarfs are transit features in the light-curve of WD\,1145+017, which are thought to be produced by dust clouds released from solid planetesimals orbiting around the white dwarf with a period of $\simeq$\,4.5\,hr \cite{vanderburgetal15-1,gaensickeetal16-1}. Searches for transiting debris around other white dwarfs have been unsuccessful \cite{faedietal11-1, belardietal16-1, vansluijs+vaneylen17-1}. White dwarfs are intrinsically faint, so transit searches are limited to a lower sky density compared to main-sequence systems. The probability of detecting transits is further limited by the narrow range of suitible orbital inclinations, and the duration of a planetesimal disruption event \cite{girvenetal11-1}.

The gaseous components of debris discs identified around a small number of white dwarfs probe the underlying physical properties of the discs. Double-peaked emission profiles are observed in a number of ionic transitions, such as the Ca\,{\textsc{ii}} 850-866\,nm triplet, indcating Keplerian rotation in a flat disc \cite{horne+marsh86-1}. Previous repeat observations of the gaseous debris disc at the white dwarf SDSS\,J122859.93+104032.9 (hereafter SDSS\,J1228+1040) have revealed long-term variability - on a time-scale of decades - in the shape of the emission lines \cite{manseretal16-1}, indicating ongoing dynamical activity in the system. 

We obtained short-cadence spectroscopy (100-140\,s) targeting the Ca\,{\textsc{ii}} triplet in SDSS\,J1228+1040 on 2017 April 20 \& 21, and again on 2018 March 19, April 10, and May 2. Our observations were conducted with the 10.4\,m Gran Telescopio Canarias (GTC, on La Palma) with the goal of searching for additional variability on the Keplerian orbital time-scales within the disc, which are of the order of hours \cite{supp_notes}. We detect coherent low-amplitude ($\simeq3$\,\%) variability in the strength and shape of the Ca\,{\textsc{ii}} triplet with a period of 123.4\,$\pm$\,0.3\,min (Figure 1), which is present in all three components of the triplet after subtracting the average emission line profile in the five nights of observations (Figure 1). Because the variability is detected in observations separated by over a year, it has been present in the disc for $\simeq$\,4400 orbital cycles. Using Kepler's third law, adopting the mass, $M$, of SDSS\,J1228+1040 as $M$\,=\,0.705\,$\pm$\,0.050\,M$_{\odot}$  (where 1\,M$_{\odot}$, the mass of the Sun, is 1.99$\times$10$^{30}$\,kg) \cite{koesteretal14-1}, the semi-major axis, $a$, of the orbit corresponding to the additional Ca\,\textsc{ii} emission is $a$\,=\,0.73\,$\pm$\,0.02\,R$_{\odot}$ (where 1\,R$_{\odot}$, the radius of the Sun, is 6.96$\times$\,10$^{8}$\,m).

The equivalent widths (EWs, a measure of the strength of the lines relative to the continuum) of the Ca\,{\textsc{ii}} triplet profiles are shown in Figure 2 along with the ratios of blue-shifted to red-shifted flux throughout the 123.4\,min period. This illustrates the variation in the overall brightness of the emission lines, and the strong asymmetry of the velocity of the additional flux. The variable emission shown in Fig.\,1 C \& F alternates (moves) from red-shifted to blue-shifted wavelengths as a function of phase. Assuming that the additional, variable emission is generated by gas in orbit around the white dwarf, this indicates that we only observe emission when the additional gas is on the far side of its orbit around the white dwarf, with respect to our line of sight. This additional emitting region is obscured, either by the disc or the region itself, when the material is traveling in front of the star, where we would otherwise observe the blue-shifted to red-shifted transition. We fitted sinusoids to both the EW and blue-to-red ratio data, finding them to be offset in phase by 0.14\,$\pm$\,0.01 cycles and 0.09\,$\pm$\,0.01 cycles in 2017 and 2018 respectively. These phase-shifts imply that the maximum EW is observed when the region emitting the additional flux is at its maximum visibility and thus furthest from us in its orbit around the white dwarf, whereas the maximum blue-shifted emission occurs up to 0.25\,cycles afterwards, once the region has orbited into the visible blue-shifted quadrant of the disc. The smoothness of the EW and blue-to-red ratio variations, along with the extent in orbital phase ($\simeq$\,0.4) of the variable emission in Figure 1, indicates that the emission region is extended in azimuth around the disc, rather than originating from a point source.

Several scenarios could plausibly explain the short-term emission detected from SDSS\,J1228+1040 (see supplementary text): (i) A low-mass companion, with Ca\,\textsc{ii} emission originating from the inner hemisphere irradiated by the white dwarf. This would naturally match the observed phase dependence \cite{maxtedetal06-1}. However radial velocity measurements rule out the the presence of any companion with mass greater than 7.3\,M$_{\textrm{J}}$ (where 1\,M$_{\textrm{J}}$, the mass of Jupiter, is 1.90$\times$\,10$^{27}$\,kg) \cite{supp_notes}, and the non-detection of hydrogen in the accretion disc excludes brown dwarfs and Jupiter-mass planets. (ii) Vortices have been invoked to explain non-axisymmetric structures detected in sub-mm observations of proto-planetary discs \cite{isellaetal13-1}. The presence of a weak magnetic field is expected to destroy any vortex that forms within a few orbital cycles. While our observations place only an upper limit to the magnetic field of the white dwarf $B<10-15$\,kG \cite{supp_notes}, the field strength required within the disc at SDSS\,J1228+1040 to render vortices unstable is $10\mu$G to 50\,mG. This field strength can be reached rapidly due to the exponential growth rate of the magnetic field in the disc \cite{supp_notes}, and we therefore rule out the presence of long-lived vortices in the disc. (iii) The photoelectric instability (PEI, \cite{klahr+lin05-1}) can possibly produce arc-shaped structures within a disc. However, these structures vary in both radial location and shape within the disc on the time-scale of months and we therefore rule out this scenario. (iv) A planetesimal orbiting in the disc and interacting with the dust could generate the detected gas (see Figure\,3). We exclude (i)-(iii) as possible scenarios, so argue that (iv) is the most plausible explanation for the coherent short-term variation detected in the Ca\,{\textsc{ii}} triplet lines at SDSS\,J1228+1040. 

The short period of the orbit around SDSS\,J1228+1040 requires any planetesimal to have a high density or sufficient internal strength to avoid being tidally disrupted by the gravity of the white dwarf. This contrasts with WD\,1145+017, where the debris fragments are detected on orbits consistent with the tidal disruption radius of a rocky asteroid \cite{vanderburgetal15-1}. Under the assumption that the body in orbit around SDSS\,J1228+1040 has no internal strength and that its spin period is tidally locked to its orbital period, we calculate the minimum density needed to resist tidal disruption on a 123.4\,min period as 39\,g\,cm$^{-3}$ for a fluid body deformed by the tidal forces \cite{supp_notes}. If we assume the body has enough internal strength to remain spherical, then the minimum density required reduces to 7.7\,g\,cm$^{-3}$, which is approximately the density of iron at 8\,g\,cm$^{-3}$ (however, the internal strength could be greater, and the density lower). We therefore conclude that the body in orbit of SDSS\,J1228+1040 needs some internal strength to avoid tidal disruption, and we calculate bounds on the planetesimal size, $s$, as $4$\,km\,$<s<600$\,km, with an uncertainty of 10\,\% in these values \cite{supp_notes}.

What is the origin of the planetesimal? It may be that the planetesimal is the differentiated iron core of a larger body that has been stripped of its crust and mantle by the tidal forces of the white dwarf. The outer layers of such a body would be less dense, and disrupt at greater semi-major axes and longer periods than the core \cite{verasetal17-2}. This disrupted material would then form a disc of dusty debris around SDSS\,J1228+1040, leaving a stripped core-like planetesimal orbiting within it. 

It remains unclear whether the variable emission originates from interactions with the dusty disc, or from irradiation of the surface of the planetesimal. Small bodies are known to interact with discs and induce variability in spatially resolved discs, such as the moon Daphnis, which produces the Keeler gap in the rings around Saturn, \cite{porcoetal05-1, tiscarenoetal07-1}. Some debris discs around main-sequence stars show evidence of gas generated after the main phase of planet formation \cite{dentetal14-1}. The origin of this non-primordial gas is uncertain, but it has been suggested that it could be generated by collisional vaporisation of dust \cite{czechowski+mann07-1}, or collisions between comets \cite{zuckerman+song12-1}. If the body is not interacting with the disc to generate the additional gas, then the planetesimal must be producing the gas. The semi-major axis of the planetesimal, $a$\,=\,0.73\,R$_{\odot}$, is close enough to the star that the surface of the body may be sublimating \cite{supp_notes}, releasing gas which contributes to the variable emission.

We hypothesise that gaseous components detected in a small number of other white dwarf debris discs \cite{gaensickeetal06-3,wilsonetal15-1} may also be generated by closely orbiting planetesimals. While sublimation of the inner edges of debris discs \cite{metzgeretal12-1}, and the break-down of 1--100\,km rocky bodies \cite{kenyon+bromley17-1}, have been proposed to explain gaseous debris discs at white dwarfs, not all metal polluted white dwarfs with high accretion rates and/or large infrared excesses host a gaseous component. The Ca\,{\textsc{ii}} triplet emission profiles from the gaseous debris disc around SDSS\,J1228+1040 have shown variability over 15\,yr of observations (\cite{manseretal16-1}, see also Fig.\,1 A \& D). This emission can be modeled as an intensity pattern, fixed in the white dwarf rest frame, that precesses with a period of $\simeq$\,27\,yr \cite{manseretal16-1}. Both the pattern and its precession are stable for orders of magnitude longer than the orbital time-scale within the disc ($\simeq$\,hours). Eight gaseous white dwarf debris discs are currently known; long-term monitoring of three of those systems has shown similar long-term variability to SDSS\,J1228+1040 \cite{wilsonetal15-1, manseretal16-2, dennihyetal18-1}.

The gaseous disc has been present at SDSS\,J1228+1040 for at least 15\,yr \cite{manseretal16-1}, implying that the planetesimal has survived in its current orbit for at least that long. A planetesimal on an eccentric orbit that precesses due to general relativity could explain the observed precession of a fixed intensity pattern. In this scenario, the planetesimal would need an eccentricity $e$\,$\simeq$\,0.54 (\cite{supp_notes}, Figure S8), bringing the periastron to 0.34\,R$_{\odot}$. An eccentric orbit is not unexpected, as the planetesimal would initially enter the tidal disruption radius at high eccentricities ($e > 0.98$) from further out in the white dwarf system \cite{verasetal14-1}. An eccentric orbit is supported by the observed precession of an asymmetric intensity pattern in the gaseous emission \cite{manseretal16-1}. Estimating the constraints on the size of a planetesimal with such a periastron results in a range of $2$\,km\,$<s<200$\,km with an uncertainty of 10\,\% in these values, smaller than previously calculated for a circular orbit. The results presented here show that planetesimals can survive in close orbits around white dwarfs, and the method applied is not dependent on the inclination of the disc.

\bibliographystyle{Science}


\noindent\textbf{Acknowledgements:} Based on observations made with the Gran Telescopio Canarias (GTC), installed in the Spanish Observatorio del Roque de los Muchachos of the Instituto de Astrof\'isica de Canarias, in the island of La Palma. This work has made use of data from the European Space Agency (ESA) mission {\it Gaia} (\url{https://www.cosmos.esa.int/gaia}), processed by the {\it Gaia} Data Processing and Analysis Consortium (DPAC, \url{https://www.cosmos.esa.int/web/gaia/dpac/consortium}). Funding for the DPAC has been provided by national institutions, in particular the institutions participating in the {\it Gaia} Multilateral Agreement. Based on observations made with ESO Telescopes at the La Silla Paranal Observatory under programme IDs: 595.C-0650.
\textbf{Funding:} This research has been carried out with telescope time awarded by the CCI International Time Programme. The research leading to these results has received funding from the European Research Council under the European Union's Seventh Framework Programme (FP/2007- 2013) / ERC Grant Agreement n. 320964 (WDTracer). T.R.M. acknowledges support from STFC (ST/P000495/1). J.D.L acknowledges the funding support of the Natural Sciences and Engineering Research Council of Canada. D.V. gratefully acknowledges the support of the STFC via an Ernest Rutherford Fellowship (grant ST/P003850/1). M.R.S. is thankful for support from Fondecyt (1141269). A.J.M. and M.B.D. acknowledge the support of KAW project grant 2014.0017. A.J.M. also acknowledges the support of VR grant 2017-04945. O.T. was partially supported by a Leverhulme Trust Research Project Grant. F.M. acknowledges support from the Royal Society Dorothy Hodgkin Fellowship. P.R-G. acknowledges support provided by the Spanish Ministry of Economy, Industry and Competitiveness through grant AYA-2017-83383-P, which is partly funded by the European Regional Development Fund of the European Union. This research was supported by the Jet Propulsion Laboratory through the California Institute of Technology postdoctoral fellowship program, under a contract with the National Aeronautics and Space Administration. 
\textbf{Author contributions:} C.J.M led the overall project. C.J.M., B.T.G., S.E., J.D.L., W.L., A.J.M, and D.J.W. contributed to the writing of the manuscript. T.R.M., F.M., and D.V, contributed to the interpretation of the results. C.J.M, M.H., and P.I. reduced the spectra obtained from VLT/UVES and GTC/OSIRIS. D.K. and O.T. produced the white dwarf models used by J.D.L. to calculate the magnetic field strength of the white dwarf. W.L. produced the photo-electric instability simulations of the disc. P.R-G., T.R.M., M.R.B., M.B.D., J.F., N.G.F., D.dM., S.G.P., A.Q., R.R., S.R., M.D.S., M.R.S., R.S., S.T., E.V., M.W., S.X., and S.P.Z. contributed to the proposals which lead to the data collection and discussion of the results.
\textbf{Competing Interests:} The authors declare there are no conflict of interests.
\textbf{Data and materials availability:} The data used in this research are available from the ESO VLT archive \cite{vlt_archive}, under proposal number 595.C-0650(G), and the GTC archive \cite{GTC_archive}, under proposal numbers GTC1--16ITP and GTC25--18A. The {\sc zeeman} software and the model shown in Fig.\,S5 are available from \url{https://sourceforge.net/projects/zeeman-f/}. The {\sc pencil code} software is provided here \url{https://github.com/pencil-code}, and the model shown in Fig.\,S6 \& S7 can be found in the directory pencil-code/samples/2d-tests/WhiteDwarfDisk, using version \#f4f2f16 of {\sc pencil code}.
\\

\noindent\textbf{Supplementary materials}\\
Materials and Methods, Supplementary text\\
Figures S1-S8\\
Tables S1-S3\\
References (38-113)


\clearpage

\centerline{\includegraphics[width=16cm]{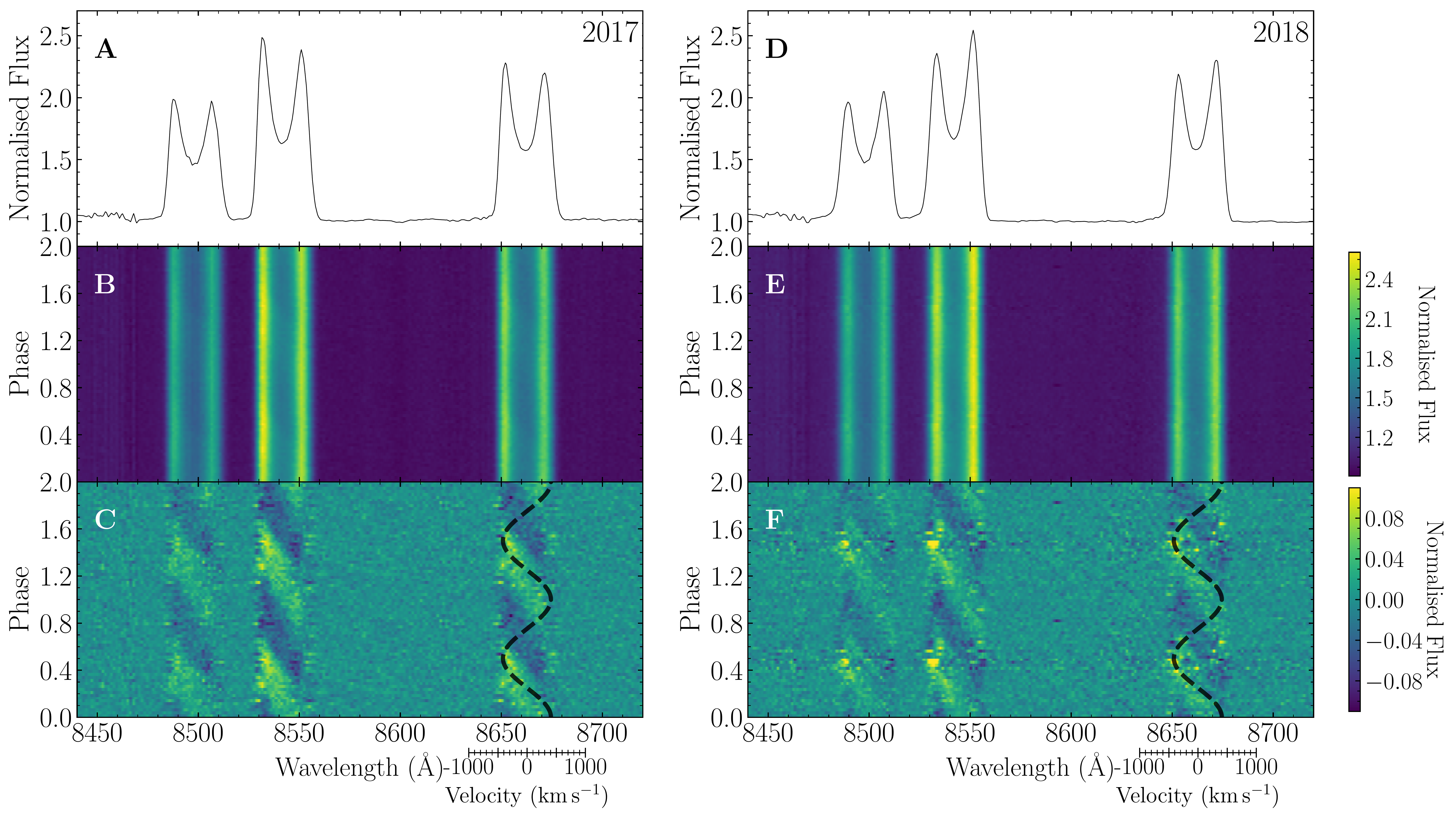}}

\medskip
\noindent {\bf Fig. 1. Phase-folded trailed spectrogram of the emission line profiles in SDSS\,J1228+1040.}
519 spectra of SDSS\,J1228+1040 were taken over two nights in 2017 (A--C), and three nights in 2018 (D--F), see Table\,S1 for a log of the observations. (A \& D) Averaged, normalised spectrum of the Ca\,{\textsc{ii}} triplet. (B \& E) Phase-folded trailed spectrograms using a 123.4\,min period (one cycle is repeated for display). The colour-map represents the normalised flux. Subtracting the coadded spectrum from the phase-folded trailed-spectrogram (done separately for each year) illustrates the variability in both flux and wavelength on the 123.4\,min period in all three components of the Ca\,\textsc{ii} triplet (C \& F). The dashed black curve is not fitted to the data, but simply illustrates the typical S-wave trail for a point source on a circular orbit with a semi-major axis of 0.73\,R$_{\odot}$ and an inclination of 73$^{\textrm{o}}$ \cite{gaensickeetal06-3}. The velocity axes refer to the longest-wavelength Ca\,{\textsc{ii}} triplet line profile.

\clearpage

\centerline{\includegraphics[width=16cm]{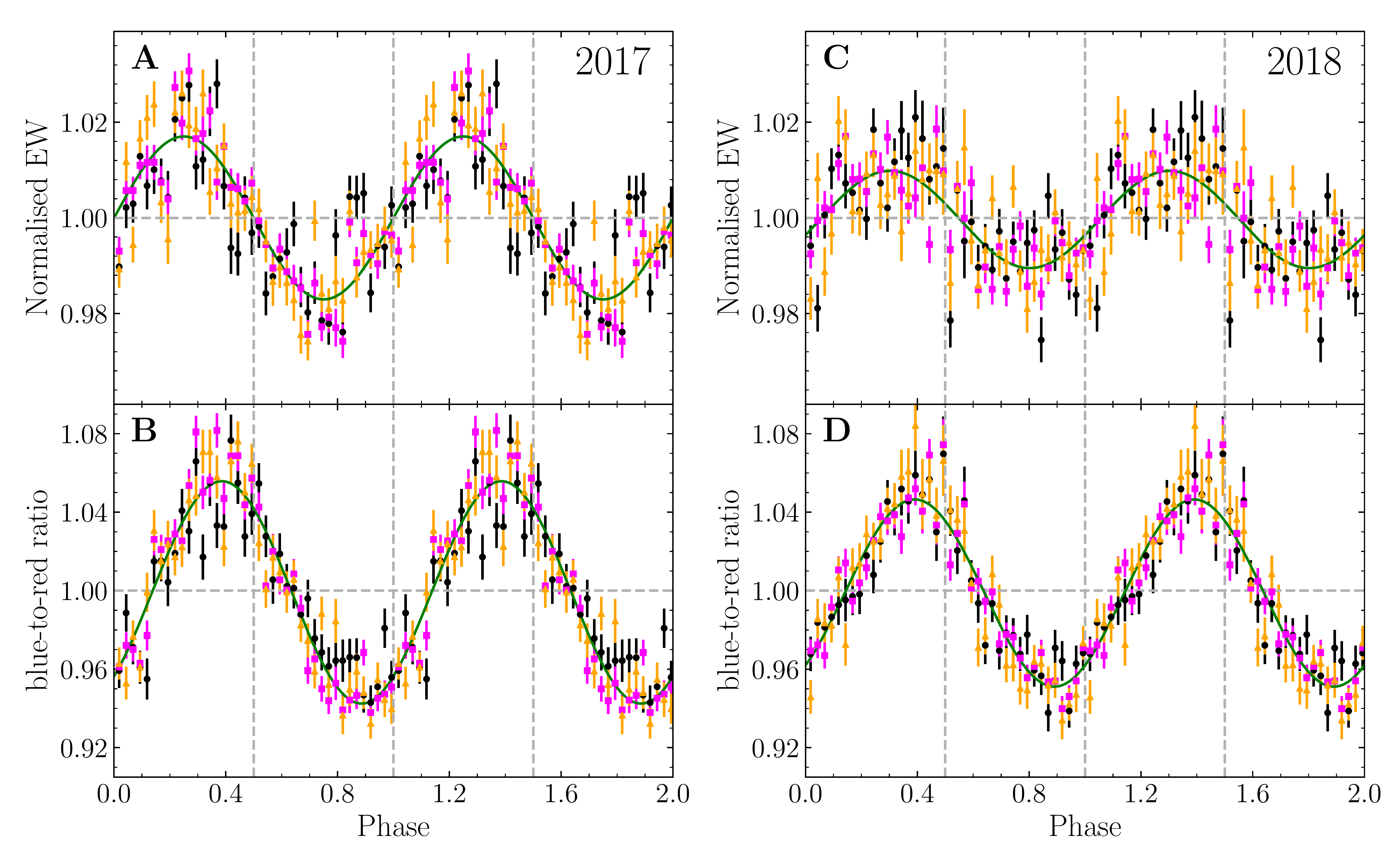}}

\medskip
\noindent {\bf Fig. 2. Variability of the Ca\,{\textsc{ii}} triplet emission of SDSS\,J1228+1040.}
Equivalent width (EW, A \& C) and blue-to-red ratio (B \& D), which is the ratio of blue-shifted to red-shifted flux centred on the air-wavelengths of the Ca\,{\textsc{ii}} triplet in the rest frame of the white dwarf at +19\,km\,s$^{-1}$, with the mean set to 1.0. The data are phase-folded on a 123.4\,min period (one cycle repeated for clarity \cite{supp_notes}) for the 2017 (A \& B) and 2018 (C \& D) data sets. The EWs and blue-to-red ratios for the 8498\,\AA, 8542\,\AA, and 8662\,\AA\ components of the Ca\,{\textsc{ii}} triplet are coloured (marked) in black (circle), magenta (square), and orange (triangle) respectively. The data are averaged over the three profiles and fitted with a sinusoid (green line). The EW and the blue-to-red ratio curves are offset in phase by 0.14\,$\pm$\,0.01 cycles (49$^{\textrm{o}}$\,$\pm$\,4$^{\textrm{o}}$) and 0.09\,$\pm$\,0.01 cycles (31$^{\textrm{o}}$\,$\pm$\,5$^{\textrm{o}}$) for the 2017 and 2018 profiles respectively. Phase zero for both the 2017 and 2018 data sets has been shifted such that the fit to the 2017 EW data passes through zero at zero phase, and the vertical dashed lines denote the phases 0.5, 1.0, and 1.5.

\clearpage

\centerline{\includegraphics[width=10cm]{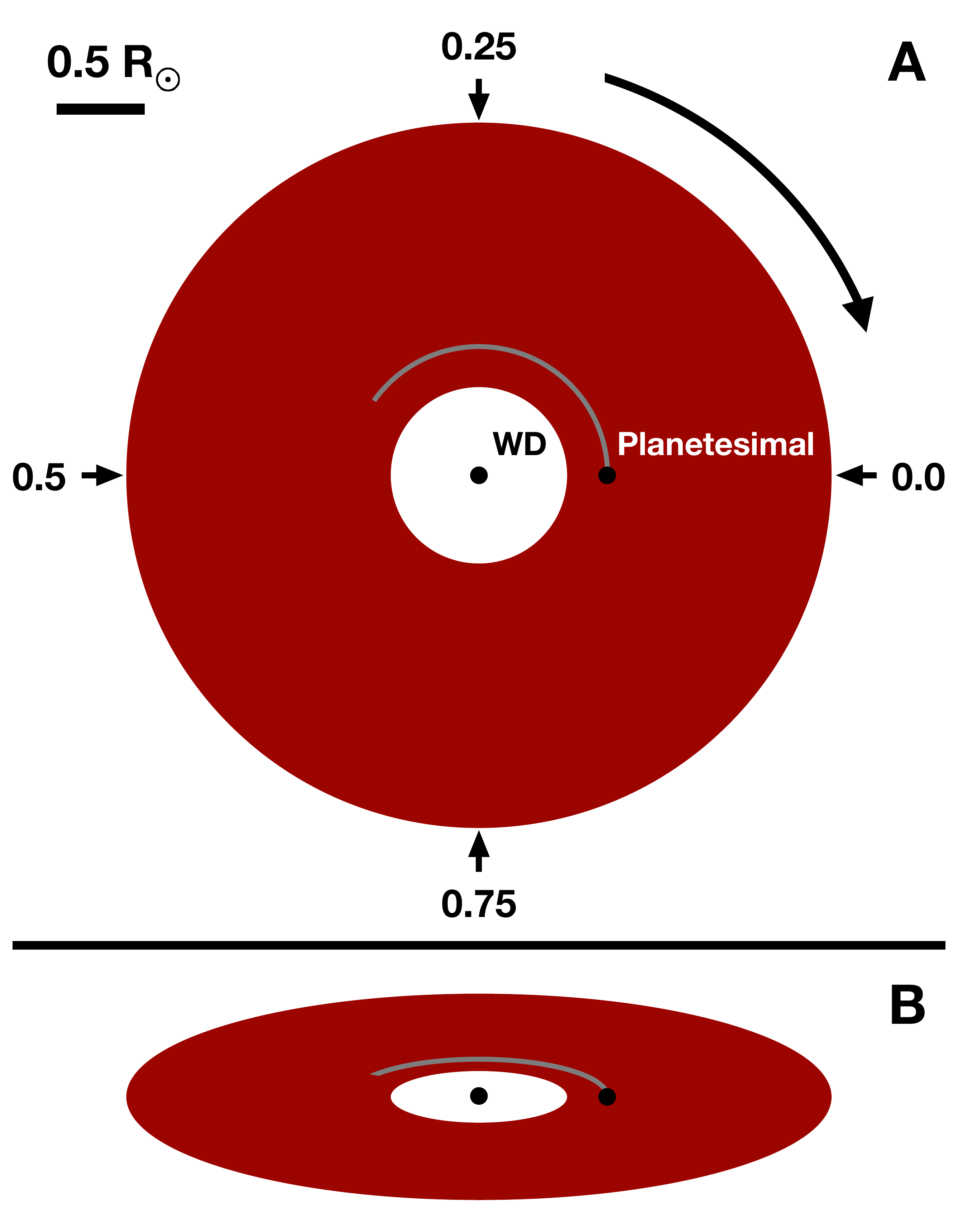}}

\medskip
\noindent {\bf Fig. 3. Schematic for the disc structure of SDSS\,J1228+1040.}
Panel A shows a top-down view of the disc around SDSS\,J1228+1040 with a planetesimal orbiting within the disc, assuming circular orbits. Both the disc and the planetesimal orbit clockwise indicated by the curved arrow, and the lines of sight for specific phases from Fig.\,1 are labelled by the straight arrows. The solid red region of the disc indicates the location of the observed Ca\,{\textsc{ii}} triplet emission, and the grey curved line trailing the planetesimal shows the azimuthal extent ($\simeq$\,0.4 in phase) of the gas stream generating the extra emission seen in Fig.\,1 C \& F. Panel B shows the system at an inclination of $i = 73^{\textrm{o}}$, as viewed from Earth \cite{gaensickeetal06-3}.

\clearpage

\setcounter{page}{1}

\centerline{\textbf{\Large Materials and Methods}}

\section{Observations}

SDSS\,J1228+1040 was observed at the 10.4\,m GTC in 2017 April 20 \& 21 and 2018 March 19, April 10, and May 2 using the OSIRIS spectrograph \cite{cepa10-1}, with the volume-phased holographic R2500I grating, and the data were obtained using 2$\times$2 pixel binning and a readout speed of 200 kHz. This setup provided a wavelength range of 733--1000\,nm with a spectral resolution $\simeq$\,0.35\,nm. We obtained a total of 519 exposures over the five nights, see Table\,S1 for full details. The GTC observations of SDSS\,J1228+1040 were reduced using standard techniques under the \textsc{Starlink} software package. The science frames were bias-subtracted and flat-fielded, and sky-subtraction and extraction of the 1-D spectra were performed using the \textsc{Pamela} software package \cite{currieetal14-1}, where the optimal-extraction algorithm was used to maximise the spectral signal-to-noise ratio. The \textsc{Molly} package \cite{molly} was used for wavelength calibration of the extracted 1-D data by coadding a set of arcs which were taken at the beginning of each night, to produce nightly HgAr + Ne + Xe arcs. Arc-lines were mapped within \textsc{Molly} and fitted with 3rd-order polynomials which were subsequently used to wavelength-calibrate the observations, and we then normalised the continuum of each spectrum with a 7th-order polynomial.

To quantify the variability detected in the Ca\,{\textsc{ii}} lines, we calculated the EW of the three individual Ca\,{\textsc{ii} triplet line profiles, as well as the strength of the blue- and red-shifted sides of the profiles (Fig.\,S1\,\&\,S2, see also Table\,S2). The EWs were calculated by integrating the flux below the line profiles in the intervals $8470-8520$\,\AA, $8524-8568$\,\AA, and $8640-8690$\,\AA, for the 8498\,\AA, 8542\,\AA, and 8662\,\AA\ emission profiles, respectively. We split the blue- and red-shifted sections for each profile using the air-wavelengths of the Ca\,{\textsc{ii}} profiles in the rest frame of the white dwarf which is at +19\,km\,s$^{-1}$ \cite{manseretal16-1}. Both the EW of the profiles, as well as the ratio of flux (blue-to-red) emitted between the blue- and red-shifted sections show variability. 

Underlying the periodic signal, there are longer-term variations affecting the EW and blue-to-red ratios, both related to the observing conditions and intrinsic to the system. The blue-to-red ratio data  points in Fig.\,S1 show a general decrease over time, which we attribute to systematic uncertainties in the continuum normalisation, which is affected by variations in airmass, as well as in the telluric absorption features that dominate either side of the Ca\,{\textsc{ii}} triplet from 7500--10000\,\AA. As such, we expect slow, systematic drifts in the measurements of the EWs and blue-to-red ratios. In addition, the nightly average EW measurements of the 2018 profiles in Fig.\,S2 (see also Table\,S2) change more than can be explained by variations in the continuum normalisation, revealing variability in the strength of the Ca\,{\textsc{ii}} triplet on a time-scale of weeks to months. The amplitude of these variations is larger than that of the two-hour signal we discuss in the main text, and cause artifacts in both the phase-folded trailed spectrogram, and the phase-folded EW and blue-to-red ratio curves. As such, we scale the strength of the 2018 EW profiles when producing Fig.\,1\,\&\,2 to that of the average strength of the 2018 March 19 data.

\section{Determination of the period of variability}
We analysed the EW and blue-to-red ratio curves using the \textsc{MIDAS/TSA} package  \cite{schwarzenberg-czerny96-1}. We combined the measurements for the three Ca\,{\textsc{ii}} components, and then computed discrete Fourier transforms for the two consecutive nights of data taken in 2017, and for the three nights of data taken several weeks apart in 2018 (Table\,S1). 

The amplitude spectra computed from the 2017 data (Fig.\,S3 A, B, \& C)} show several possible period aliases separated by 1\,d$^{-1}$, as it is usually the case for single-site data. We fitted sine functions to the time-series measurements to determine the uncertainties of the periods corresponding to the three strongest aliases (Table\,S3). To evaluate the likelihood of the individual aliases representing the true period of the short-term variability we ran a bootstrap simulation (\cite{press02-1}, their chapter 15.6) and found that the most likely periods (and their probabilities) measured from the variability of the equivalent widths and the blue-to-red ratios are 122.88$\pm$0.19\,min (98.6\%) and 123.63$\pm$0.15\,min (98.0\%). These two periods are consistent at the $3\sigma$ level, and folding the Ca\,{\textsc{ii}} profiles on either of them results in equally smooth trailed spectrograms. We attribute the small discrepancy between the two period measurements to the systematic differences in the morphology of the time-series data of the equivalent width and blue-to-red ratio, and the fact that only two to three phase cycles were obtained during each of the two nights.

The 160 spectra obtained in 2018 were spaced out in three observing runs separated by $\simeq$3 weeks each, corresponding to several hundred cycles of the Ca\,{\textsc{ii}} variability. Given that these three sets of data only span $\simeq$0.85 to 1.79 phase cycles each, the individual sets provides a period measure with an accuracy of $\simeq5$\%~--~which is insufficient to derive a unique period from the combined 2018 observations. The amplitude spectrum computed from the equivalent widths is less well defined than that computed from the blue-to-red ratios, which we attribute to the variation in the nightly average of the overall Ca\,{\textsc{ii}} equivalent width. As an initial test, we simply folded the 2018 Ca\,{\textsc{ii}} profiles on either of the 2017 periods, which results in trailed spectrograms that are very similar to that obtained from the 2017 data.

We computed amplitude spectra from the combined 2017 and 2018 blue-to-red ratio data, which results in strong one-day aliases superimposed with a very fine high-frequency alias structure from the week-long and year-long gaps in the time-series. The best-fitting period from this data set is $P=123.4\pm0.3$\,min, which we adopt for further analysis. The uncertainty was set to reflect the difference between the two periods derived above from the 2017 data. The Ca\,\textsc{ii} profiles for the 2017 and 2018 observations were phase-folded on this period to produce Fig.\,1. We rescaled the average equivalent width of each of the spectra obtained on 2018 April 10 and 2018 May 2 to that measured from the 2018 March 19 profiles before phase folding to remove artifacts generated by the long-term variations in equivalent width. An example spectrum for phase 0.4875 from the 2017 spectra is shown in Fig.\,S4. 

We are confident that we have identified the one-day alias corresponding to the true period of the Ca\,{\textsc{ii}} variability. While adopting the period corresponding to either neighbouring alias changes the numerical results, our general conclusions remain unaffected from the choice of the alias. For example, if we adopt the neighbouring periods to our best fitting value, $P=114.04$\,min, and $P=135.01$\,min, the semi-major axis of the orbit changes to 0.69\,R$_{\odot}$ and 0.77\,R$_{\odot}$ respectively, and the upper limit on a planetesimal size calculated below, changes to 550\,km and 650\,km respectively, a difference of $\simeq$\,10\,\%.

\section{Parameters of the white dwarf SDSS\,J1228+1040}

\subsection{Distance and Mass}

Most parameters derived for a planetesimal in orbit around SDSS\,J1228+1040 (period, semi-major axis, eccentricity, size, tidal heating) depend on the mass of the white dwarf, which has been measured \cite{koesteretal14-1} to be $M$\,=\,0.705\,$\pm$\,0.050\,M$_{\odot}$. Using a mass-radius relation, the distance to the system has been estimated using photometry in the optical and UV, as 120.9\,$\pm$\,9.4\,parsec and 134.2\,$\pm$\,9.9\,parsec, respectively \cite{koesteretal14-1}. The \textit{Gaia} Data Release~2 \cite{gaia16-1, gaia18-1} reports a parallax of 7.89\,$\pm$\,0.09\,milliarcseconds for SDSS\,J1228+1040 (Gaia source\_id\,=\,3904415787947492096), corresponding to a distance of 126.7\,$\pm$\,1.5\,parsec, which is consistent with the two distance estimates \cite{koesteretal14-1}, and we therefore adopt that mass.

\subsection{Magnetic field strength}\label{s-mag}

The non-detection of Zeeman splitting in the Balmer lines of SDSS\,J1228+1040 rules out magnetic fields $B\ge1$\,MG, which are detected in $\simeq2-5$\% of white dwarfs \cite{kepler13-1, kleinmanetal13-1, sionetal14-1}. The incidence of weaker fields among white dwarfs is still poorly constrained \cite{landstreetetal12-1}, however, fields of 70--500\,kG have been detected from the splitting of metal lines in a significant fraction (three out of a sample of fourteen) of cool white dwarfs of spectral type DAZ \cite{kawka+vennes14-1}, the same spectral type as SDSS\,J1228+1040. If such a field were present in SDSS\,J1228+1040 it would affect the accretion process from the disc into the stellar atmosphere, and possibly affect the planetesimal. To derive an upper limit on the field strength in SDSS\,J1228+1040, we have compared the observed photospheric metal lines with model spectra of magnetic white dwarfs.

We computed synthetic line profiles of the line triplets of Si\,{\sc ii} at 4128--4130\,\AA, and of Mg\,{\sc ii} at 4481\,\AA\ and compared those to a high-resolution spectrum of SDSS\,J1228+1040 obtained on 2017 March 01 with UVES on the VLT \cite{deekeretal00-1}, which was reduced using the {\sc reflex} reduction work flow \cite{freudlingetal13-1} with standard settings. Si\,{\sc ii} and Mg\,{\sc ii} profiles were computed for a mean field modulus $\langle | B | \rangle$ (i.e. the average value of the field modulus over the observable hemisphere) ranging from zero to 50\,kG. The computations were carried out using the {\sc fortran} code {\sc zeeman} \cite{landstreet88-1, bailey+landstreet13-1}. This code requires a model atmosphere structure appropriate for the atmospheric parameters of SDSS\,J1228+1040 \cite{somgaensickeetal12-1}, which was computed with the code of \cite{koester10-1}. {\sc Zeeman} solves the LTE radiative transfer problem of radiation emerging locally, as modified by a specified magnetic field, for spectral line profiles in all four Stokes parameters. The local emergent line profiles are then summed over the visible stellar disc, appropriately Doppler shifted to account for stellar radial velocity and rotation, to produce a predicted set of (Stokes $I$) line profiles for the spectral region being studied. For SDSS\,J1228+1040 a dipolar field configuration, with the factor-of-two contrast between polar and equatorial field strengths of a pure dipole somewhat reduced, was assumed.

Examples of computed profiles are compared to the observed lines in Fig.\,S5. It is clear, particularly from the very sharp Si\,{\sc ii} lines, that in order to escape detection, a field in SDSS\,J1228+1040 would have to have $\langle | B | \rangle \le10-15$\,kG, a field strength close to the weakest field detected in any white dwarf \cite{landstreetetal17-1}.

\centerline{\textbf{\Large Supplementary Text}}

\setcounter{section}{0}

\section{Alternative scenarios causing the observed Ca\,\textsc{ii} variability}

\subsection{Stellar/Sub-stellar companion}

We consider the possibility of a stellar or sub-stellar (brown dwarf or Jovian planet) in orbit around SDSS\,J1228+1040, where the observed variability in the Ca\,{\textsc{ii}} triplet emission profile could be explained by the irradiated inner-hemisphere of a companion. Such emission line variability has been detected in H~$\alpha$ in SDSS\,J1557+0916, arising from the accretion of hydrogen from a 63\,M$_{J}$ brown dwarf in a 136.4\,min orbit with a white dwarf \cite{somfarihietal17-1}. This system also contains a dusty debris disc polluting the white dwarf photosphere with metals.

Radial velocity measurements of the Mg\,{\textsc{ii}} 4481\,\AA\ line put a limit on the mass, $M_{\textrm{p}}$ for any possible companion to SDSS\,J1228+1040 at $M_{\textrm{p}}\sin{i}$\,$\leq$\,7\,M$_{J}$, and adopting an inclination of 73$^{\textrm{o}}$ obtained from modelling the Ca\,{\textsc{ii}} emission line profiles \cite{gaensickeetal06-3}, we obtain an upper limit on the companion mass of $M_{\textrm{p}}$\,$\leq$\,7.3\,M$_{J}$. The spectrum of SDSS\,J1228+1040 lacks the emission of hydrogen detected in all white dwarf plus brown dwarf binaries \cite{maxtedetal06-1, littlefairetal14-1, casewelletal18-1}, which is also seen in cataclysmic variables in which white dwarfs accrete from low-mass main-sequence stars \cite{manser+gaensicke14-1}. We therefore rule out the presence of a hydrogen-rich stellar or sub-stellar companion. Another possible analogue are AM\,CVn stars, a small class of binaries containing white dwarfs accreting from hydrogen-depleted degenerate companions, some of them with extreme mass ratios \cite{kupferetal16-1}. However, all AM\,CVn stars exhibit strong emission lines of helium, which are also not detected in the spectrum of SDSS\,J1228+1040. The material transferred to white dwarfs in Cataclysmic Variables (CVs) - binary systems containing a white dwarf accreting from a hydrogen-rich donors - and AM\,CVn stars (with hydrogen-depleted donors) is rich in carbon and nitrogen, respectively, both of which are strongly depleted in the material accreted by SDSS\,J1228+1040 \cite{somgaensickeetal12-1}. The absence of hydrogen and helium emission lines, together with the fact that the abundances of the material accreted onto the white dwarf in SDSS\,J1228+1040 are compatible with a rocky parent body, rules out the presence of any type of stellar or sub-stellar companion filling, or close to filling, their Roche lobe. Finally, assuming a typical radius of a brown dwarf at $\simeq$\,1\,R$_{\textrm{J}}$ (where 1\,R$_{\textrm{J}}$, the radius of Jupiter, is 6.99$\times$\,10$^{7}$\,m) \cite{laughlin18-1}, we calculate a minimum mass required for a companion to not fill its Roche lobe (using Kepler's third law and the radius at which a companion would share the same volume as its Roche lobe, \cite{warner95-1} their equation 2.3b) as $\simeq$\,18\,M$_{\textrm{J}}$, which is a factor two larger than our upper limit mass estimate. We therefore exclude a brown dwarf or a jovian planet as possible explanations for the Ca\,\textsc{ii} variability detected at SDSS\,J1228+1040. We are also able to exclude large planetary bodies, see below.

\subsection{A vortex in the disc}\label{s-vortex}

Dust trapping vortices have been invoked to explain non-axisymmetric structures in sub-mm observations of protoplanetary discs \cite{isellaetal13-1, vandermareletal13-1, casassusetal13-1, marinoetal15-1}. The origin of these structures and their conclusive identification as vortices have not yet been determined from observations, but theoretical analyses and numerical simulations have determined some of the main properties of disc vortices. One of the most robust routes for their origin is the Rossby wave instability (RWI, \cite{lovelace+hohlfeld78-1, lovelaceetal99-1, lietal00-1}), which is triggered by a 20\,\%-30\,\% localized axisymmetric increase in pressure, with the extra shear converted into vorticity. In numerical simulations of primordial protoplanetary discs the RWI is pervasive because this condition is easily realised at the boundaries between turbulent and quiescent zones \cite{varniere+tagger06-1, lyraetal09-2, lyra+maclow12-1}, at planetary gaps \cite{koller03-1, devalborroetal07-1, lyraetal09-1}, and transitions in resistivity/viscosity \cite{lyraetal15-1, flocketal17-1}. 

Vortices are known to be destroyed by the magnetoelliptic instability (MEI, \cite{mizerski+bajer09-1, lyra+klahr11-1, faureetal15-1}), a weak (subthermal) field instability that is a generalised form of the magnetorotational instability (MRI, \cite{balbus+hawley98-1}) in flows with elliptical streamlines \cite{mizerski+lyra12-1}. The MEI is a weak-field instability and should be present if the conditions for the MRI are also present. We can assess the conditions for the MRI in the disc around SDSS\,J1228+1040. 

Adopting an upper limit on the field strength of 10\,kG for the white dwarf (see Section\,\ref{s-mag}), the implied upper limit on the field strength in disc - due to the decrease in magnetic field strength with radius, $r$ as $B \propto r^{-3}$ - is 10-100\,mG. At a temperature, $T$\,=\,6000\,K and a column density, $\Sigma$\,=\,10$^{-4}$\,g\,cm$^{-2}$ \cite{kinnear11, melisetal10-1}, the ratio of thermal to magnetic pressure in the disc $\beta$ is $\sim$\,20\,--\,2000 for a field strength of 100\,mG and 10\,mG respectively, and the weak-field condition is therefore satisfied. The ionisation fraction is also high, so the gas around SDSS\,J1228+1040 should be MRI-active. The growth rate, of MEI is exponential, and the amplification, $r_{\textrm{amp}}$, of seed instabilities can be calculated as

\begin{equation}\tag{S1}
r_{\textrm{amp}} = e^{0.75\frac{2\pi}{P}t},
\end{equation}

\noindent
where $t$ is the length of time, and over three orbits ($t$\,=\,$3P$), the amplification of MEI is $\sim$\,10$^{6}$ \cite{lyra+klahr11-1, mizerski+lyra12-1}. From this growth rate, we can thus conclude that any vortex in this gas will be highly unstable to the MEI and quickly destroyed.

In the analysis above, we assumed a two-dimensional (2D) disc. The aspect ratio, $h$ of the disc is $\simeq$\,0.005, assuming a disc radius of $\simeq$\,0.73\,R$_{\odot}$, and a disc scale height $H$\,$\simeq$\,4.3\,$\times$\,10$^{-3}$\,R$_{\odot}$ (\cite{melisetal10-1}, their equation 12) with a stellar mass $M_{*}$\,=\,0.705\,M$_{\odot}$, and a distance from the star $r$\,=\,0.73\,R$_{\odot}$. From this, we conclude that the disc is flat and can be treated as 2D. As MEI is a version of MRI, it will be suppressed if the MRI wavelength, $\lambda_{\rm MRI}$ is greater than the scale height of the disc, $\lambda_{\rm MRI} > H$. We can rewrite this to determine when the MRI stabilizes, which becomes a condition for the plasma $\beta$,

\begin{align}
H < \lambda_{\rm MRI}, \tag{S2}\\
H < 2\pi v_A/\Omega, \tag{S3}\\
c_s < 2\pi v_A, \tag{S4}\\
c_s^2/v_A^2 < 4\pi^2, \tag{S5}\\
\beta < 8  \pi^2  ~ 80, \tag{S6}
\end{align}

\noindent where $v_{A}$ is the Alfv\'en speed, $\Omega$ is the differential rotation of the disc, and for $h=0.005$, the sound speed, $c_s \simeq$\,2\,km\,s$^{-1}$. This sets the upper limit to the magnetic field in the disc at 50\,mG, compatible with the upper limit derived for the field strength of the white dwarf.
This is self-consistent as the MRI and the MEI are weak-field instabilities. We therefore rule out the hypothesis that a vortex in the disc is generating the Ca\,\textsc{ii} variability detected in the high-cadence GTC spectroscopy of SDSS\,J1228+1040.

In principle, it is possible to estimate the accretion rate, $\dot{M}_{\textrm{acc}}$, onto the white dwarf due to MRI turbulence using

\begin{equation}\label{e-accrate}\tag{S7}
\dot{M}_{\textrm{acc}} = \frac{3 \pi \Sigma \alpha c^{2}_{s}}{\omega},
\end{equation}

\noindent where $\alpha$ is the viscosity parameter, $\omega$\,=\,$2\pi/P$ is the angular frequency, and $c_{s}$ is the sound speed given by $c_{s}$\,=\,$T c_{p} (\gamma - 1)$, where $T$ is the temperature of the disc, $c_{p}$ is the heat capacity at constant pressure, and $\gamma$ is the adiabatic index \cite{shakura+sunyaev73-1, lynden-bell+pringle74-1}. Estimates of the column density span many orders of magnitude, from $\Sigma$\,$\sim$\,10$^{-9}$ to 0.3\,g\,cm$^{-2}$ \cite{kinnear11, melisetal10-1, hartmannetal11-1}. The lower limit of 10$^{-9}$\,g\,cm$^{-2}$ is determined by the fact that no emission from forbidden line cooling is detected in the spectrum of SDSS\,J1228+1040 \cite{melisetal10-1}. The upper limit of 0.3\,g\,cm$^{-2}$ assumes that the disc is viscously heated to produce the Ca\,{\sc ii} emission \cite{hartmannetal11-1}, and results in an inferred accretion rate of $\sim$\,10$^{17}$\,g\,s$^{-1}$, which is many orders of magnitudes above the highest accretion rate observed in any debris-accreting white dwarf \cite{bergforsetal14-1}. We consider this upper limit as totally physically unrealistic. Adopting, as an example, $\Sigma$\,=\,10$^{-4}$\,g\,cm$^{-2}$, and $T$\,=\,6000\,K, which was determined from a photo-ionisation model for the  Ca\,{\textsc{ii}} triplet emission \cite{kinnear11}, $P$\,=\,123.4\,min, and $\alpha$\,=\,0.25 \cite{wilsonetal14-1}, we estimate an accretion of $\dot{M}_{\textrm{acc}}$\,=\,4\,$\times$\,10$^{10}$g\,s$^{-1}$. We conclude that for a value of $\Sigma\simeq10^{-6}$\,g\,cm$^{-2}$, consistent with the current estimates of the column density, MRI turbulence would result in an accretion rate that is broadly consistent with the accretion rate derived from modelling the photospheric metal abundances, 5.6\,$\times$\,10$^{8}$g\,s$^{-1}$ \cite{somgaensickeetal12-1}.

Self-gravity in the disc is negligible, considering the Toomre parameter, $Z = c_s \Omega / (\pi G \Sigma)\simeq9\times10^{12}$, where $G$ is the gravitational constant. Self gravity is only relevant for $Z < 1$. 

\subsection{Photoelectric instability}\label{s-PEI}

Another possibility for the origin of the brightness asymmetry in the gas disc at SDSS\,J1228+1040 is the photoelectric instability (PEI, \cite{klahr+lin05-1, klahr+lin01-1, besla+wu07-1, lyra+kuchner13-1}), which operates in a cycle: (i) electrons are ejected off dust grains by ionising radiation, (ii) the superthermal electrons heat the gas via collisions, (iii) dust grains move toward the high pressure gas, (iv) more dust leads to more photoelectric ionisations, releasing more heat, and a further increase in the dust concentration, resulting in a positive feedback. Models of the photoelectric instability in 2D and 3D find that it produces rings and arcs in both gas and dust \cite{lyra+kuchner13-1}. Including radiation pressure in the model produces a variety of other structures, including spirals and large eddies \cite{richertetal18-1}.

Although the PEI was originally proposed for gaseous debris discs around pre-main-sequence and main-sequence stars \cite{klahr+lin05-1}, the process should occur in any optically thin disc of gas and dust illuminated by a photo-ionising source. Systems with a high gas-to-dust mass ratio ($\sim$\,1) should experience PEI, though in dust-dominated discs the PEI should also be present, albeit only in nonlinear form (\cite{lyra+kuchner13-1}, their figure 1). 

Our two sets of time-resolved spectroscopy show that the two-hour variability in the Ca\,{\textsc{ii}} line is present in SDSS\,J1228+1040 over at least 4000 orbits, whereas simulations of PEI only extend to 400 orbits \cite{richertetal18-1}. To assess if the photoelectric instability could result in structures in the disc that are sufficiently long-lived to explain the observed variability, we simulated a disc for 2000 orbits, and scrutinised the time evolution of these structures. The low bolometric luminosity of white dwarfs renders radiation pressure unimportant, so the model of \cite{lyra+kuchner13-1} which we apply here is more applicable than that of \cite{richertetal18-1}. 

The PEI model is calculated in cylindrical coordinates, in 2D (see above) in the disc midplane, with radial range $r=[0.4,2.5]R_{\odot}$ and full $2\pi$ coverage in azimuth. The number of cells, $L_r$ and $L_\phi$ for the radius and azimuth respectively is $L_r \times L_\phi = 256\times 256$. We added 500\,000 Lagrangian particles to this grid to represent the dust component. Dust and gas interact through drag forces, and we do not include relativistic effects. The equations of motion are \cite{lyra+kuchner13-1}:

\begin{align}
\ptderiv{\varSigma_g} &=& -\left(\v{u}\cdot\del\right) \varSigma_g -\varSigma_g \Div{\v{u}}, \tag{S8}\\
\ptderiv{\v{u}} &=& -\left(\v{u}\cdot\del\right) \v{u} - \frac{1}{\varSigma_g}\grad{P} - \grad{\varPhi} -\frac{\varSigma_d}{\varSigma_g}f_d, \tag{S9}\\
P &=& \cv \left(\gamma-1\right)T_0 \varSigma_0^{-1} \varSigma_g\varSigma_d  + \varSigma_g c_b^2, \label{eq:pressure} \tag{S10}\\
\frac{d\v{v}}{dt} &=& -\grad{\varPhi} + f_d, \tag{S11}\\
f_d &=& -\frac{\left(\v{v}-\v{u}\right)}{\tau_f}. \tag{S12}
\end{align}

\noindent where $\varSigma_g$ and $\varSigma_d$ are the gas and dust surface density, respectively, $\v{u}$ and $\v{v}$ are the gas and dust velocities. $P$ is the gas pressure, $\varPhi$ is the gravitational potential of the white dwarf, $\tau_f$ is the timescale of aerodynamical drag between gas and dust, $\gamma=1.4$ is the adiabatic index, and $\cv=c_{P}/{\gamma}$ is the specific heat capacity at constant volume, where $c_P = 1$ is the specific heat capacity at constant pressure. In the equation of state (Equation\,\ref{eq:pressure}), the first term embodies a simple prescription for photoelectric heating in the instantaneous thermal coupling approximation \cite{lyra+kuchner13-1}, to avoid having to solve the energy equation. The second term represents a basal pressure set by other heating sources. For simplicity we fix the basal sound speed $c_b^2 = \Theta c_s^2$ with $\Theta \equiv {\rm const} = 0.5$.

The initial condition is a disc without a global pressure gradient using

\begin{align}
c_s(r,\phi) = const = c_{s,0} = 0.1, \tag{S13}\\
\rho(r, \phi) = const = \rho_{0} = 1.0, \tag{S14}
\end{align}

\noindent
given in code units where $c_{s,0} = 2.25$\,km\,s$^{-1}$ and $\rho_{0} = 1.5 \times 10^{-13}$\,g\,cm$^{-3}$, to prevent aerodynamical dust drift. The radial boundary condition is zero radial velocity in the inner boundary and outflow in the outer boundary, linear extrapolation in logarithm for the azimuthal velocity and density. At the boundaries - the inner one up to $r=0.5$\,R$_{\odot}$, and the outer one down to $r=2.35$\,R$_{\odot}$ - the quantities are driven back to their initial condition, within a time $t=0.1 T_0$ where $T_0$ is the orbital period at the reference radius $r=1 R_{\odot}$. Sixth-order hyper-dissipation terms are added to the evolution equations to provide extra dissipation near the grid scale \cite{lyraetal17-1}. These terms are needed for numerical stability because the high-order scheme of the {\sc Pencil Code} \cite{pencil_code} has little overall numerical dissipation \cite{mcnallyetal12-1}. They are chosen to produce Reynolds numbers of order unity at the grid scale, but then drop as the sixth power of the scale at larger scales, so they have negligible influence on the large-scale flow. Shock diffusion is added to the equations of motion, to resolve shocks to a differentiable length \cite{richertetal15-1, lyraetal16-1, hordetal17-1}. Extra Laplacian viscosity is added to the equations, with $\alpha= 10^{-2}$ \cite{shakura+sunyaev73-1}.

An equal number particles by area are randomly distributed over the disc, with velocities set to their Keplerian value. Particles are placed between $r=[0.5,2.4]R_{\odot}$ and removed from the domain if they cross these boundaries. The dust grains have Stokes number $\St=\tau_f\varOmega = 1$, and we start them with dust-to-gas ratio $\varepsilon \equiv \varSigma_p / \varSigma_g =1$. These values were chosen to maximise the growth rate of PEI in the debris disc. The backreaction of the drag force is added to the gas, conserving momentum in the system.

Panels A \& B of Fig.\,S6 show, at every radii, the azimuthal average of the dust and gas densities vs time respectively, while panels C \& D show the dust and gas density respectively at the end of the simulation. The dust quickly rearranges into a series of regularly spaced arcs, which are labelled a--f in Panel A. We are primarily interested in checking whether any structure in the disc can remain constant in shape and location over 1000s of orbits. Rings form at $r\approx$\,0.7, 1.25, and 1.75 $R_{\odot}$, but they are evanescent and soon disperse. A longer lived one at $r=$\,1.75\,$R_{\odot}$ is sustained until about 700 orbits. After that, the arc systems remain for long timescales. The system e eventually disperses at 1750 orbits, leaving systems a--d and f until the end of the simulation. The systems b--e all moved outwards during the course of the calculation, driven by the pressure gradient they effect on the gas. The systems a and f are more stable spatially, due to boundary conditions imposed on the computations. For either a to move inwards and f to move outwards, crossing the domain boundaries, they would need to climb pressure gradients. The fact that only the arc systems affected by boundaries retain their integrity over long timescales is a good indication that the structures produced by photoelectric instability are unlikely the origin of the variable emission we detect on a 123.4\,min period.

In Fig.\,S7 we shows the azimuthal power spectrum of the dust density as a function of time, broken down by azimuthal wavenumber to illustrate the time evolution of the azimuthal substructure in the same arc systems shown in Fig.\,S6. Most of the power is in the $m=0$ mode; the number of arcs is shown as the dominant azimuthal wavenumber below the $m=0$ line. As seen, the arc systems in panels A--E alternate between $m=1$ (one arc) and $m=2$, some with $m=3$, and no structure is long lived. The arc system in F is still growing in intensity at the end of the simulation and did not achieve a steady state.

In summary, we conclude that arcs of gas generated by PEI cannot account for for the azimuthally extended emission around SDSS\,J1228+1040. 

Given that the three scenarios outlined above are all unlikely to cause the short-term variability of the Ca\,\textsc{ii} emission lines detected in SDSS\,J1228+1040, we conclude that a solid body orbiting with a semi-major axis $a=0.73\,R_\odot$ is the most plausible hypothesis.

\section{Constraints on the size of the planetesimal}\label{s-body_size}

We estimate lower and upper limits on the size (see \cite{sombrownetal17-1}, their section 3.2), mass and lifetime of a planetesimal orbiting SDSS\,J1228+1040 with a semi-major axis, $a$\,=\,0.73\,\,R$_{\odot}$ using two different assumptions: (i) The accretion rate onto the white dwarf, $\dot{M}_\mathrm{WD}$ = 5.6$\times$10$^8$\,g\,s${^{-1}}$ \cite{somgaensickeetal12-1}, is generated entirely from the sublimation of the planetesimal, which would be the dominant source of gas in the system. (ii) The maximum size a body with binding forces dominated by internal strength can be before being tidally disrupted. For both scenarios we assume a core-like composition, e.g. iron dominated.

\subsection{A lower limit from the measured accretion rate}\label{s-acc}

Under the assumption that the accretion rate onto the white dwarf is equal to the rate of sublimation, $\dot{M}_\mathrm{sub}$ of a planetesimal, we calculated the size, $s$, of the object as

\begin{equation}\label{e-sub}\tag{S15}
s = \frac{a}{R_{\textrm{WD}}{T^2}_{\textrm{WD}}} \left(\frac{L_{\textrm{vap}}{\dot{M}_{\textrm{sub}}}}{\sigma} \right)^{0.5},
\end{equation}

\noindent
where $R_{\textrm{WD}}$\,=\,0.01169\,R$_{\odot}$ and ${T}_{\textrm{WD}}$\,=\,20713\,K are the radius and temperature of the white dwarf respectively \cite{manseretal16-1}, $L_{\textrm{vap}}$\,=\,6.09$\times$10$^{4}$\,J\,g$^{-1}$ is the latent heat of vaporisation for iron \cite{somdean99-1}, and $\sigma$ is the Stefan-Boltzmann constant \cite{sombrownetal17-1}. This gives $s$\,$\simeq$\,4\,km, which we take as a lower limit to the size of the planetesimal, as any shielding from the debris disc would decrease the received radiation, and would require an increase in size to match the rate of accretion onto the white dwarf. A body of this size would have a mass $M_{\textrm{p}}$\,$\simeq$\,10$^{18}$\,g and a lifetime, $t$\,$\simeq$\,85\,yr.

While these calculations assume that the planetesimal has a pure iron composition, the accreted material in the atmosphere of the white dwarf is similar to bulk Earth \cite{somgaensickeetal12-1}. However, it is possible that the solid body is the surviving core of a larger planetesimal, where the outer layers have formed the debris disc and are currently accreting onto the white dwarf. We also calculate the lower limit on size and lifetime under the assumption that the body is rocky, with a latent heat of vaporisation, $L_{\textrm{vap}} = 8 \times 10^3$\,J\,g$^{-1}$ \cite{sombrownetal17-1}, as $s$\,$\simeq$\,1\,km and $t$\,$\simeq$\,1.5\,yr, respectively. This should be considered as a strict lower limit, as the corresponding life time only slightly exceeds the time span over which the short-term Ca\,{\textsc{ii}} variability has been detected.

\subsection{An upper limit from internal strength considerations}\label{s-str}

We discussed in the main text that an orbit with a semi-major axis of $a$\,=\,0.73\,R$_{\odot}$ requires that the planetesimal has some amount of internal strength, and we can hence calculate size limits for the planetesimal assuming the forces opposing tidal disruption are dominated by the internal strength. Using a range of internal strengths $S$\,=\,40--1000\,MPa obtained from iron meteorite and iron-nickel samples \cite{sompetrovic01-1, somslyuta13-1}, we calculate the maximum size an iron-dominated planetesimal can reach before it is tidally disrupted using (\cite{sombrownetal17-1}, their equation 29),

\begin{equation}\label{e-tid}\tag{S16}
s = \left(\frac{2Sa^{3}}{GM_{\textrm{WD}}\rho_{\textrm{iron}}}\right)^{0.5},
\end{equation}

\noindent
where $M_{\textrm{WD}}$\,=\,0.705\,M$_{\odot}$ is the mass of the white dwarf \cite{manseretal16-1}, and $\rho_{\textrm{iron}}$\,$\simeq$\,8\,g\,cm$^{-3}$ is the density of iron \cite{sombrownetal17-1}. This gives a range of sizes $s$\,$\simeq$\,60\,--\,600\,km, which corresponds to a mass range of 7\,$\times$\,10$^{21}$\,--\,7\,$\times$\,10$^{24}$\,g. The upper end of this range is comparable to the largest asteroid in the Solar System, Ceres, with a radius of 473\,km \cite{somrusselletal16-1}. Calculating the sublimation rate for bodies in this size range by rearranging Equation\,\ref{e-sub} results in $\dot{M}_\mathrm{sub}$\,$\simeq$\,10$^{11}$\,--\,10$^{13}$\,gs$^{-1}$, and a lifetime range of 1\,400\,--\,14\,000\,yr respectively. These sublimation rates are higher than the measured accretion rate onto the white dwarf, but shielding from the disc may reduce this.

In summary, we have shown that a planetesimal (which could be as large as Ceres) can survive within the disc for a minimum of about a century and possibly far longer. It is likely that the body is shielded somewhat by the radiation from the white dwarf.

\section{An eccentric orbit}\label{s-eccorb}

We speculate that the short-term variability identified in the GTC observations is linked to the long term variability detected in the gaseous emission at SDSS\,J1228+1040. The morphology of the Ca\,{\textsc{ii}} triplet emission profiles varies on a period of $\simeq$\,27\,yr \cite{manseretal16-1}, and we suggest that a planetesimal that is subject to general relativistic precession could interact with the dust in the debris disc, generating the gaseous disc component and establishing the intensity pattern in that disc. The fixed intensity pattern observed in the gaseous disc at SDSS\,J1228+1040 is asymmetric, and should smear out on a few orbital time-scales, unless it is maintained in some way, such as by a planetesimal on an eccentric orbit. We calculate an eccentricity of the intensity pattern at SDSS\,J1228+1040 as $e \simeq 0.4$, under the assumption that the minimum and maximum velocities observed in the Doppler map \cite{manseretal16-1}, are at the apastron and periastron of the orbits respectively.

Modeling of the evolution of hydrodynamical eccentric discs has shown that the observed, long-term variability could be explained by the precession of a disc due to general relativistic precession or pressure in the disc \cite{miranda+rafikov18-1}. The precession period in this scenario has a strong dependence on the radius of the inner edge of the gas disc, and can explain the range of precession periods observed so far \cite{manseretal16-1,dennihyetal18-1}. However, it is unknown whether the observed inner edge is due to a change in the density in the disc, or a change in the ionisation. We suggest that a planetesimal on an eccentric orbit could induce both the eccentricity and range of precession timescales seen in the gaseous debris discs observed so far.

Fig.\,S8 shows that for the body orbiting at 0.73\,R$_{\odot}$ to precess at the same rate as the fixed intensity pattern in the disc, it must have an eccentricity of $e$\,$\simeq$\,0.54. This brings the closest approach of the planetesimal to the white dwarf to $a_{\textrm{c}}$\,=\,$a(1-e)$\,=\,0.34\,R$_{\odot}$, where $a$ is the semi-major axis. Bodies on an eccentric orbit are not tidally locked to their host star, but will enter a state of pseudosynchronisation where their spin angular velocity is approximately the orbital angular velocity at pericentre. $e$\,=\,0.54 implies a spin period for the body in orbit around SDSS\,J1228+1040 of $\simeq$\,40\,mins (\cite{hut81-1}, their equation 42), which is roughly three times faster than the orbital period of 123.4\,mins. The more rapid spin rate will increase the minimum density required to avoid tidal disruption. We recalculate this density as 13\,g\,cm$^{-3}$ (\cite{verasetal17-2}, their equation 5), with a modified constant $\mathcal{R}$\,$\simeq$\,3.75 determined by re-balancing the increased tidal force with the centripetal force, against gravitational force from the planetesimal (see also \cite{murray+dermott99-1}, their section 4.8). Our original conclusion that the body requires internal strength still holds.

Using the closer approach of 0.34\,R$_{\odot}$ to the white dwarf, we re-evaluate the constraints on the size of the planetesimal as 2\,km and 20\,--\,200\,km respectively. Both of these calculations assume that the body spends all of its time at a distance of 0.34\,R$_{\odot}$ from the white dwarf, so the size range calculated here is a conservative estimate.

Below we calculate the heating and eccentric decay of a body due to induction and tidal heating. We calculate these values using the planetesimal radius $r_{\textrm{p}}$, which is related to the size definition of the planetesimal by $s$\,$\simeq$\,1.65\,$r_{\textrm{p}}$ (see \cite{sombrownetal17-1}). As such, our lower and upper bounds to the radius of the planetesimal are $r_{\textrm{p}}$\,=\,1.2--120\,km.

\subsection{Eccentricity damping timescale}

We estimate here the time taken for a planetesimal on an eccentric orbit to decay onto a circular orbit using a linear, constant $Q$ tidal model, where $Q$ is a parameter that incorporates the fraction of tidal energy dissipated per orbit (usually in the range 1-100 for rocky bodies). We calculate the timescale, $\tau_{e}$, for the eccentricity to dampen for this constant $Q$ tidal model as:

\begin{equation}\label{e-eccdis}\tag{S17}
\tau_{e} = \frac{e}{\dot{e}} = \frac{4}{63} 
\left(\frac{M_{\textrm{p}}}{M_{\textrm{WD}}}\right)
\left(\frac{a}{r_{\textrm{p}}}\right)^{5}
\left(\frac{\tilde{\mu}Q}{n}\right),
\end{equation}

\noindent
where $r_{\textrm{p}}$ is the radius of the planetesimal, $n$ is the mean motion given by $\frac{2\pi}{P}$, and $\tilde{\mu}$ is the ratio of elastic to gravitational forces given by:

\begin{equation}\tag{S18}
\tilde{\mu} = \frac{19\mu}{2 \rho g_{\textrm{p}} r_{\textrm{p}}},
\end{equation}

\noindent
where $\mu$ is the modulus of rigidity and is $\sim$\,50\,GPa for iron or rock \cite{rayne+chandrasekhar61-1}, and, $g_{\textrm{p}}$ is the surface gravity of the planetesimal \cite{murray+dermott99-1}. Substituting the range in $r_{\textrm{p}}$ as 1.2-120\,km, and using $Q$\,=\,10 from above, we obtain a range $2$\,Myr\,$<\tau_{e} < 2\times$\,10$^{8}$\,Myr for the eccentricity to decay. This is orders of magnitude longer than the time-scale of the observations as well as the estimates for the lifetime of the planetesimal, and confirms the plausibility of a planetesimal on an eccentric orbit.
\\

\subsubsection{Tidal heating}

The damping of an eccentric orbit due to tidal forces will also heat up the orbiting body. We calculate the power of this heating, $P_{\textrm{T}}$, as:

\begin{equation}\label{e-eccheat}\tag{S19}
P_{\textrm{T}} = \frac{63}{4}
\left(\frac{e^2 n}{\tilde{\mu}Q}\right)
\left(\frac{r_{\textrm{p}}}{a}\right)^{5}
\left(\frac{GM_{\textrm{WD}}^2}{a}\right),
\end{equation}

\noindent
and for planetesimals with radii in the range 1.2--120\,km, we obtain a heating due to tidal forces of 4\,$\times$\,10$^{2}$--4\,$\times$\,10$^{16}$\,W \cite{murray+dermott99-1}. There is a strong dependence on the size of the planetesimal, and we compare these values for tidal heating to the heat received due to the radiation, $P_{\textrm{R}}$, from the white dwarf as,

\begin{equation}\label{e-radheat}\tag{S20}
P_{\textrm{R}} \simeq \pi \sigma_{\textrm{sb}} T_{\textrm{eff}}^4
\left(\frac{R_{\textrm{WD}} r_{\textrm{p}}}{a}\right)^2,
\end{equation}

\noindent
where $\sigma_{\textrm{sb}}$ is the Stefan-Boltzmann constant. For the planetesimal size range above, we obtain the range in radiation power received as $\simeq$\,10$^{13}$--10$^{17}$\,W. It is clear that tidal heating is not important for planetesimal sizes close to the lower limit, however, for sizes close to the upper limit, $r_{\textrm{p}}$\,=\,120\,km, tidal heating contributes at most $\simeq$\,40\,\% more than produced from the maximum radiation heating alone. Assuming tidal and radiation heating are in equilibrium with the re-radiated flux from the planetesimal, this would lead to a $\simeq$\,10\,\% increase in the temperature of the body, which is unlikely to affect its long-term evolution on its close orbit around the white dwarf.

\subsection{Induction heating}

In order for magnetic induction to act, the planetesimal has to experience non-negligible anisotropic magnetic flux. This can be caused by an anti-alignment of the stellar magnetic field with the orbital plane of the planet \cite{kislyakovaetal17-1}. Currently the relative inclination of the magnetic field and rotation axis of SDSS\,J1228+1040 are unconstrained, and hence magnetic induction heating of the planetesimal cannot be excluded.

We determined above the upper limit on the magnetic field strength of the white dwarf at SDSS\,J1228+1040 as 10--15\,kG. Assuming a dipole configuration, we calculate a conservative upper limit on the magnetic field strength at the orbital distance of the planetesimal as $B_{0}$\,$\sim$\,0.1\,G (1\,$\times$\,10$^{-5}$\,T), assuming the field strength at the white dwarf is $\simeq$\,25\,kG.

The eddy current density, $\textbf{J}$, induced in a homogeneous spherically symmetric conductive sphere moving through an asymmetric magnetic field with a frequency identical to the mean motion, $n$, can be calculated via

\begin{equation}\tag{S21}
\textbf{J}(r,\theta) = \hat{\phi}
\left(\frac{-3 i n \sigma_{\textrm{con}} B_{0} r_{\textrm{p}}}
{2 k r_{\textrm{p}} j_{1}'(k r_{\textrm{p}}) + 4j_{1}(k r_{\textrm{p}})}\right) j_{1}(kr)\sin{\theta},
\end{equation}

\noindent
where $\hat{\phi}$ is the unit vector in the azimuthal direction, $\theta$ is the angle of elevation, $r$ is the radial distance measured from the center of the planetesimal, $i^2 = -1$, $\sigma_{\textrm{con}}$ is the conductivity of the material, $B_{0}$ is the magnetic flux at the surface of the planetesimal, $j_{1}$ is the Bessel function of the first kind and $j_{1}'$ is its derivative in the radial direction. Furthermore, $k$ is the complex wavenumber, $k$\,=\,$(-i n \sigma_{\textrm{con}} \mu_{0})^{1/2}$ \cite{nagel18-1}, where $\mu_{0}$ is the vacuum permeability. The penetration (skin) depth, $\delta$, of the magnetic field in the conducting sphere is given by $\delta = (2/(n \sigma_{\textrm{con}} \mu_{0}))^{1/2}$, which gives the radial distance from the surface at which the field strength has dropped by a factor of $\simeq$\,$e=2.718$.

The power dissipated in the body due to induction heating can be calculated using Ohm's law, given by,

\begin{equation}\tag{S22}
P_{\textrm{I}} = \int \sigma_{\textrm{con}}^{-1} \textbf{J} \textbf{J*} dV,
\end{equation}

\noindent
and by substituting in $\textbf{J}$ the induction heating $P_{\textrm{I}}$ is given by

\begin{equation}\label{e-indheat}\tag{S23}
P_{\textrm{I}} = -\frac{3 \pi}{2}
\left(\frac{B_{0}}{\mu_{0}}\right)^{2}
\frac{r_{\textrm{p}}}{\sigma_{\textrm{con}}}
\left(2 + d r_{\textrm{p}}
\frac{\sin(d r_{\textrm{p}}) + \sinh(d r_{\textrm{p}})}
{\cos(d r_{\textrm{p}}) - \cosh(d r_{\textrm{p}})}\right),
\end{equation}

\noindent
where $d = 2 / \delta$. Assuming a conductivity of iron rich materials at temperatures close to the radiative equilibrium temperature ($T_{\textrm{eq}}$\,$\simeq$\,2200\,K, see Section\,\ref{s-temp}) as $\sigma_{\textrm{con}}$\,$\simeq$\,2\,MS\,m$^{-1}$ \cite{pozzoetal14-1}, we obtain $\delta$\,$\simeq$30\,m. In the range of $r_{\textrm{p}}$\,=\,1.2--120\,km and $\sigma_{\textrm{con}}$\,$\simeq$\,2\,MS\,m$^{-1}$, $d r_{\textrm{p}} >> 1$, and Equation\,\ref{e-indheat} can be reduced to

\begin{equation}\label{e-indheat_simple}\tag{S24}
P_{\textrm{I}} \simeq 3 \pi
\left(\frac{B_{0}}{\mu_{0}}\right)^{2}
\frac{r_{\textrm{p}}^{2}}{\sigma_{\textrm{con}} \delta}.
\end{equation}

\noindent
For a range of planetesimal radii 1.2--\,120\,km, and using $B_{0}$\,=\,1\,$\times$\,10$^{-5}$\,T, $\sigma_{\textrm{con}}$\,=\,2\,MS\,m$^{-1}$, and $\delta$\,=\,30\,m, we calculate the heating due to magnetic induction as 10$^{1}$--10$^{5}$\,W. This is about 12 orders of magnitude weaker than the maximum heating of the planetesimal by irradiation from the white dwarf, and as such we neglect heating and orbital decay generated by the induction heating.\\

\section{Temperature estimate of the planetesimal}\label{s-temp}

Under the assumption that the planetesimal is tidally locked to the star, we can estimate the substellar and the mean temperature over the hemisphere of the planetesimal facing the white dwarf using

\begin{equation}\tag{S25}
T_{\textrm{eq,ss}} \simeq \sqrt{\frac{R_{\textrm{WD}}}{a}} T_{\textrm{eff}},
\end{equation}

\noindent
and,

\begin{equation}\tag{S26}
\langle T_{\textrm{eq}} \rangle \simeq 2^{-\frac{1}{4}} \sqrt{\frac{R_{\textrm{WD}}}{a}} T_{\textrm{eff}},
\end{equation}

\noindent
respectively, where $T_{\textrm{eff}}$ is the effective temperature of the white dwarf, $a$ is the semi-major axis of the planetesimal's orbit, and $R_{\textrm{WD}}$ is the radius of the white dwarf. Given $T_{\textrm{eff}}$\,=\,20713\,K, and $R_{\textrm{WD}}$\,=\,0.01169\,R$_{\odot}$ \cite{manseretal16-1}, we calculate $T_{\textrm{eq,ss}}$\,$\simeq$\,2600\,K and $\langle T_{\textrm{eq}} \rangle$\,$\simeq$\,2200\,K. These values are in excess of the canonical sublimation temperature of the dusty discs at white dwarfs, which is thought to be $\simeq$\,2000\,K \cite{rafikov+garmilla12-1}. If the body is on an circular orbit and not tidally locked, then the temperature of the planetesimal will be

\begin{equation}\tag{S27}
\langle T_{\textrm{eq}} \rangle \simeq \sqrt{\frac{R_{\textrm{WD}}}{2a}} T_{\textrm{eff}},
\end{equation}

\noindent
which results in $\langle T_{\textrm{eq}} \rangle$\,$\simeq$\,1800\,K, within the expected range for the sublimation temperature of the debris disc, but higher than that of iron at $\simeq$\,1600\,K \cite{rafikov+garmilla12-1}. However, The values calculated here are estimated upper limits, and a non-zero albedo or shielding from material in the disc could reduce the temperature of the planetesimal further. For a body on an orbit with semi-major axis $a$, the time-averaged temperature of the body will decrease with eccentricity, $e$ \cite{mendez+rivera17-1}. For $e$\,=\,0.54, the decrease in the time-averaged temperature of the planetesimal is only $\simeq$\,2\,\%, so is negligible in these calculations.

\clearpage 
\centerline{\includegraphics[width=15cm]{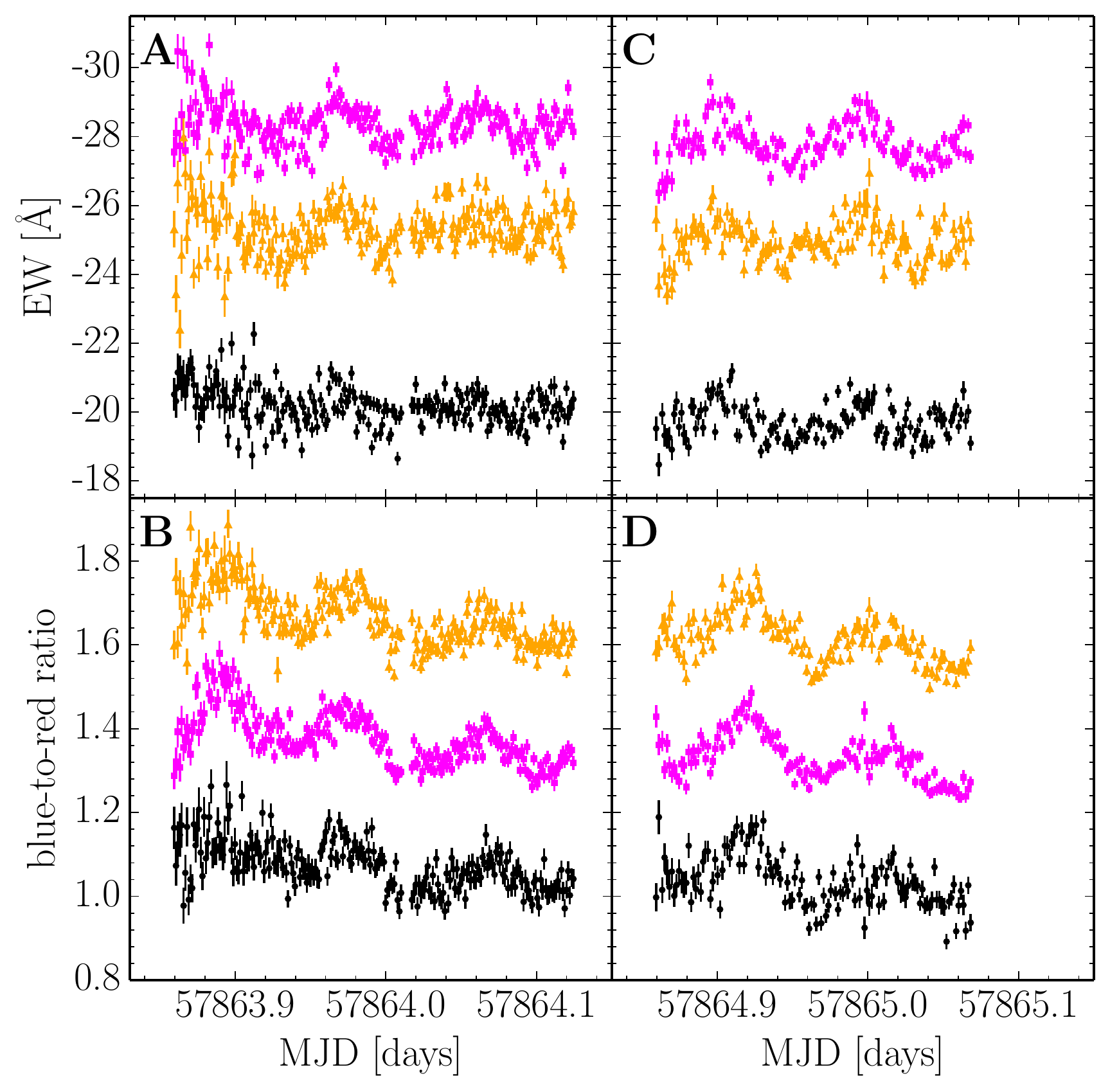}}

\medskip
\noindent {\bf Fig. S1. Equivalent width (A \& C) and blue-to-red ratio (B \& D) measurements for the Ca\,{\textsc{ii}} triplet in 2017.} The equivalent width (EW) of the Ca\,{\textsc{ii}} triplet measured from the observations taken on 2017 April 20 (A) \& 21 (C) are shown separately as a function of Modified Julian Date (MJD) for the 8498\,\AA, 8542\,\AA, and 8662\,\AA\ components of the Ca\,{\textsc{ii}} triplet, coloured (marked) in black (circle), magenta (square), and orange (triangle) respectively. The EW and blue-to-red ratio curves are both continuum normalised. The blue-to-red ratio curves are offset in steps of 0.3 from the 8498\,\AA component of the Ca\,{\textsc{ii} triplet}.

\clearpage 
\centerline{\includegraphics[width=15cm]{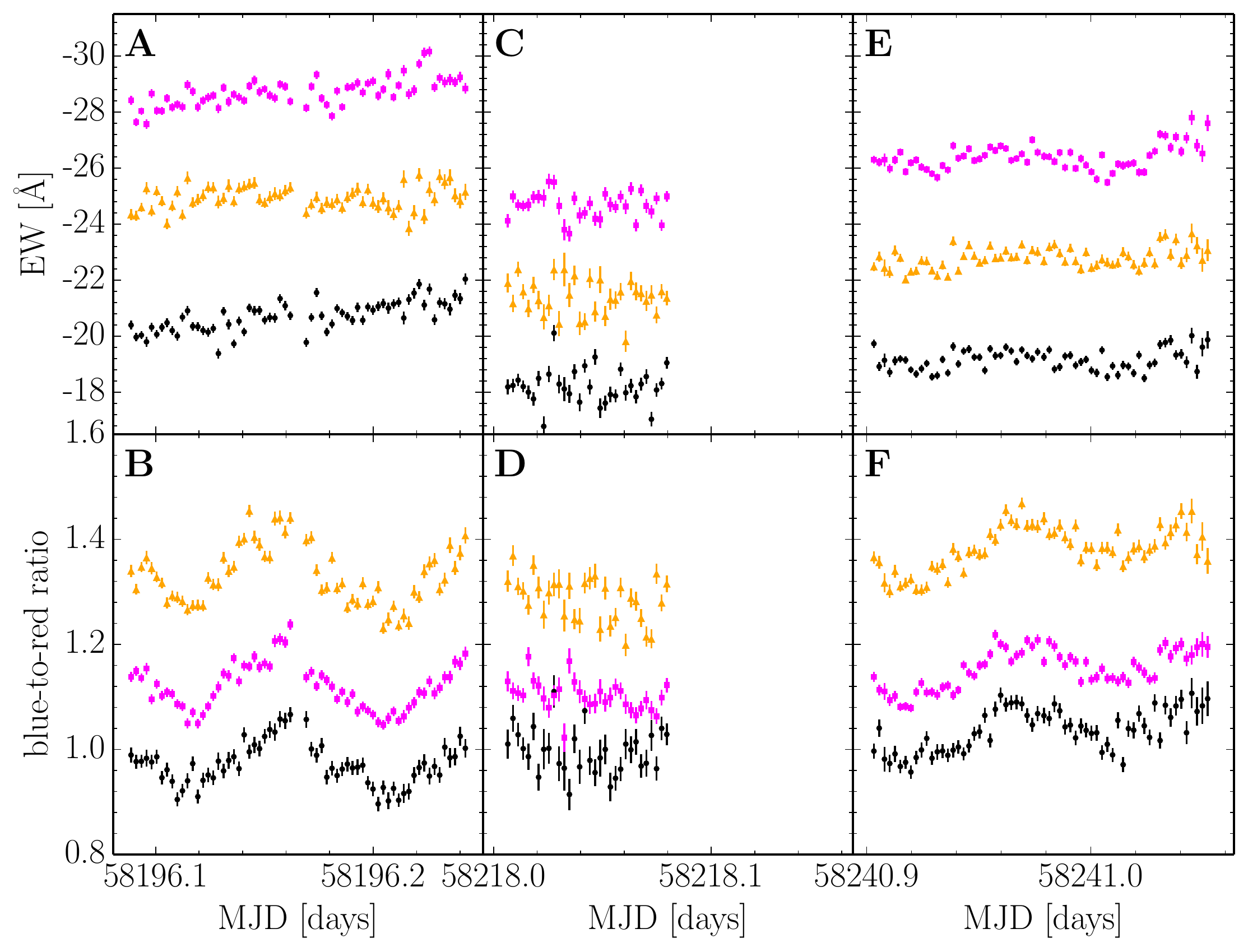}}
\medskip
\noindent {\bf Fig. S2.} Same as Figure S1, but for the 2018 observations.

\clearpage

\centerline{\includegraphics[width=16cm]{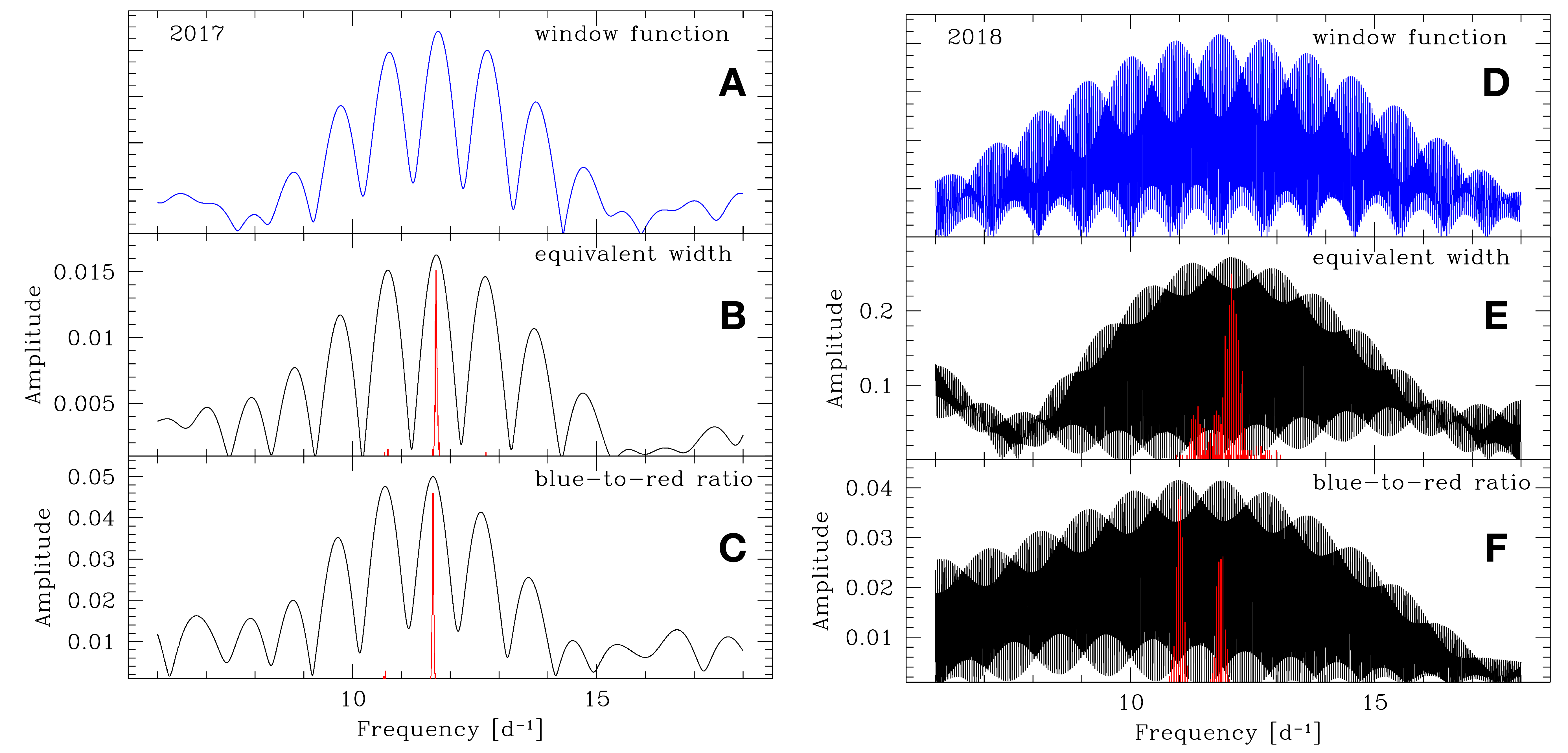}}

\medskip
\noindent {\bf Fig. S3. Amplitude spectra of the equivalent width and blue-to-red ratio measurements.} Discrete Fourier transforms (black) were computed from the normalised equivalent widths (B \& E) and blue-to-red ratios (C \& F) measured from the time-series spectroscopy obtained over two consecutive nights in 2017 (A, B, \& C) and over three nights spread over several weeks in 2018 (D, E, \& F). The amplitude spectra contain the typical 1\,$d^{-1}$ alias structure for single-site time series data, which are illustrated by the window function (blue, A \& C). In 2018, this pattern is superimposed by a fine-structure of aliases related to the long gaps inbetween the three individual observing runs (Table\,S1). The probability distribution across the different aliases was assessed using a bootstrap test, and are shown in red. Based on the larger amplitude of the blue-to-red ratio signal, and the nearly three-sigma probability (96.8\%), we identify the most likely period, which we compute as a weighted average of the values derived from the equivalent widths and blue-to-red ratios, $P=123.4\pm0.3$\,min.

\clearpage

\centerline{\includegraphics[width=11cm]{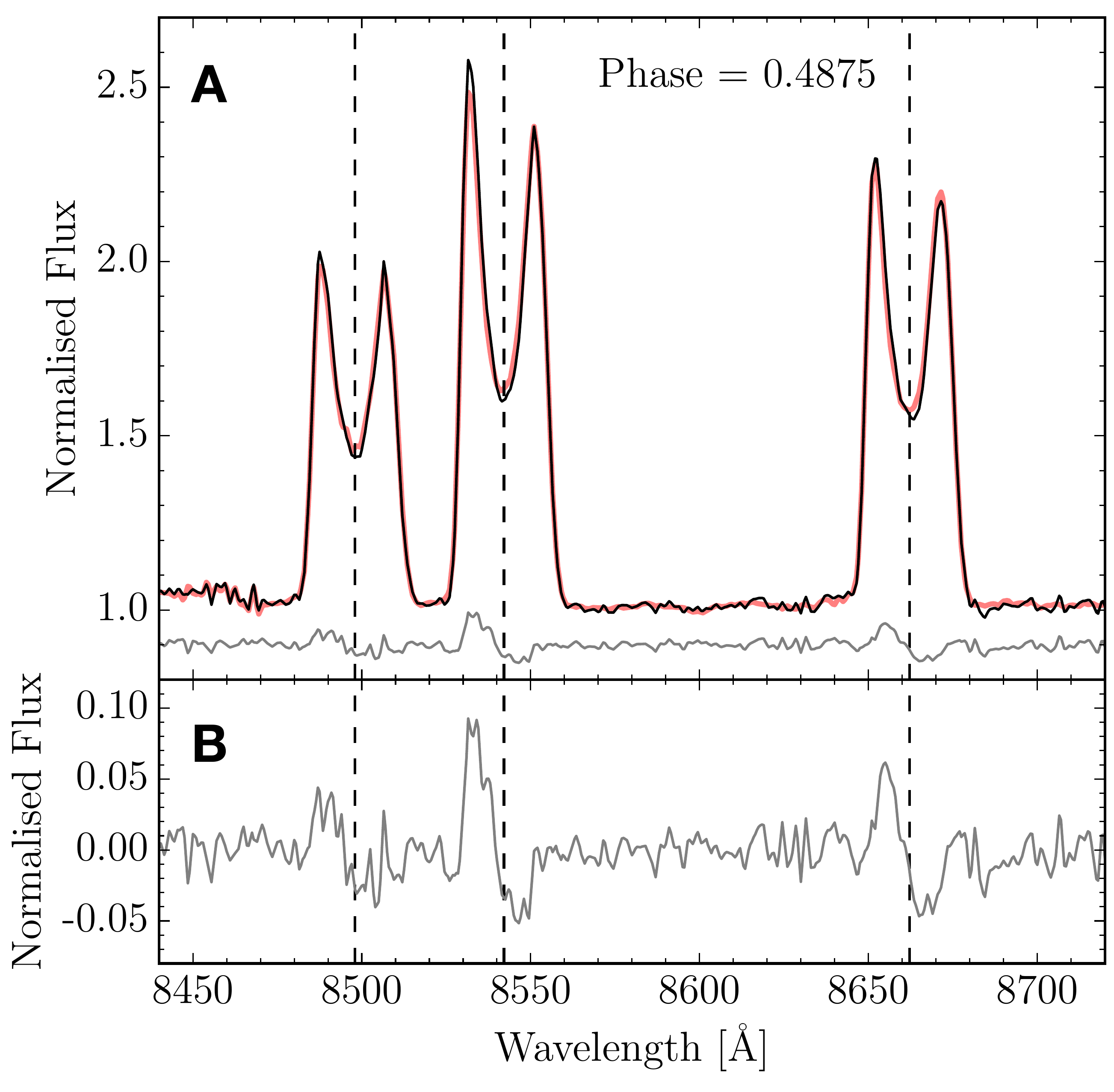}}

\medskip
\noindent {\bf Fig. S4. An example average-subtracted spectrum at phase 0.4875.} A shows the average spectrum (red) for the Ca\,{\textsc{ii}} triplet and for the phase 0.4875 (black) from the 2017 spectra (Fig.\,1 A). Subtracting the average spectrum from the averaged spectra at this phase results in the average-subtracted spectrum (grey), which has been shifted up by 0.9 for clarity. This average-subtracted spectrum is enlarged in B. The dashed lines indicate the rest wavelengths of the Ca\,{\textsc{ii}} triplet.

\clearpage
\centerline{\includegraphics[width=15cm]{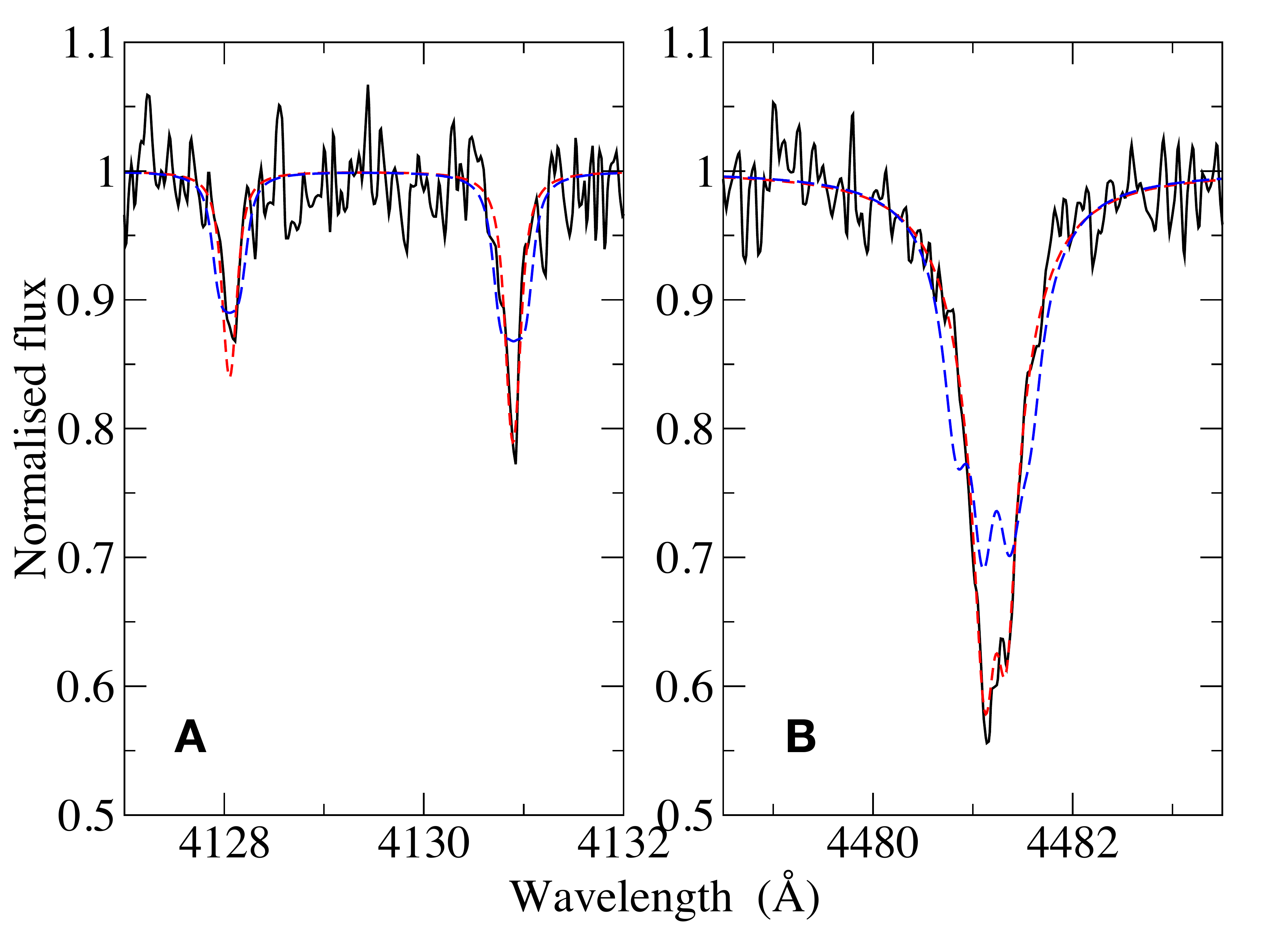}}

\medskip
\noindent {\bf Fig. S5. Illustration of the upper limit on the magnetic field strength of SDSS\,J1228+1040.} Comparison of the observed line profiles (black) of Si~{\sc ii} 4128--4131~\AA\ (A) and Mg~{\sc ii} 4481\,\AA\ (B) with computed profiles for the non-magnetic case, $\langle | B | \rangle = 0$\,kG, (red, short dash) and with profiles for which the presence of a field is visible (blue, long dash). For the Mg~{\sc ii} lines the magnetic line fit is unacceptable with a field of $\langle | B | \rangle = 29$\,kG. A weaker field strength limit can be obtained from the Si~{\sc ii} lines, which are much narrower than the blend of the two strong components of the Mg\,{\sc ii} line. The fit to the Si~{\sc ii} lines is already very poor for a weaker field of 14.5\,kG. We conclude from the Si lines that the upper limit on the field strength at SDSS\,J1228+1040 is $\langle | B | \rangle$\,=\,10--15\,kG.

\clearpage
\centerline{\includegraphics[width=15cm]{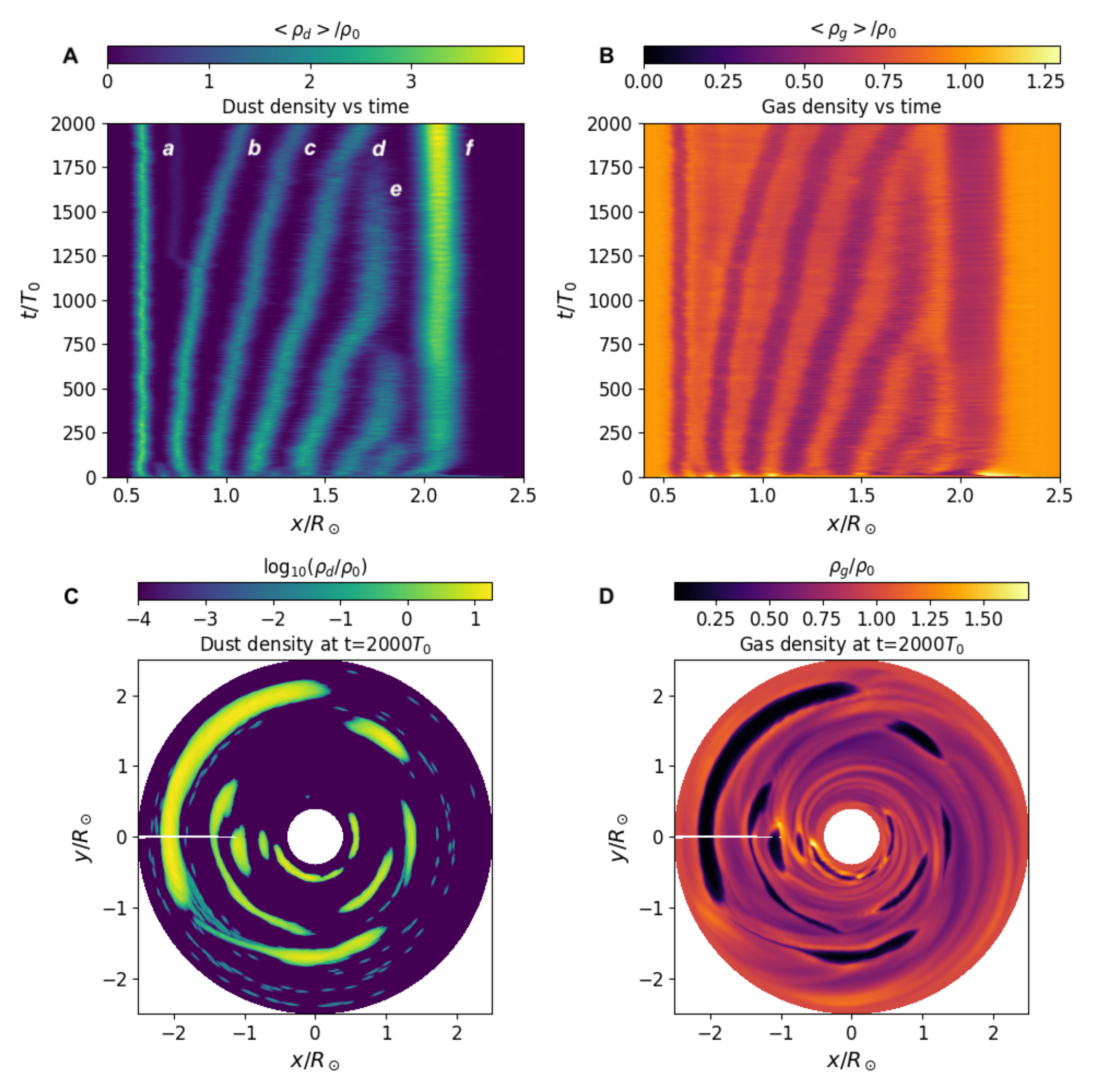}}

\medskip
\noindent {\bf Fig. S6. PEI model of the debris disc at SDSS\,J1228+1040.} A is the azimuthally averaged dust density vs radius and time. Some of the structure is evanescent, while five arc systems, a--d, and f, survive to the end of the simulation. Of these, the middle three drift outwards. System e survived for a long time, but dispersed at around 1750 orbits. B shows the azimuthally averaged gas density vs radius and time. As the dust heats up the gas, the gas expands and hence the gas and density structures are anti-correlated. Hot gas with lower densities is located where the dust density is high, and cold dense gas is found where the dust density is low. C shows the dust density distribution at the end of the simulation at $t=2000$ orbits. Azimuthal substructure is apparent as some orbits have more than one arc. D shows the gas density distribution at the end of the simulation at $t=2000$ orbits. The white lines at y=0, x$<$0 in the bottom panels are a numerical artifact as the simulation extends from -3.1415 to 3.1415 rads, not exactly from $-\pi$ to $\pi$ rads.

\clearpage
\centerline{\includegraphics[width=15cm]{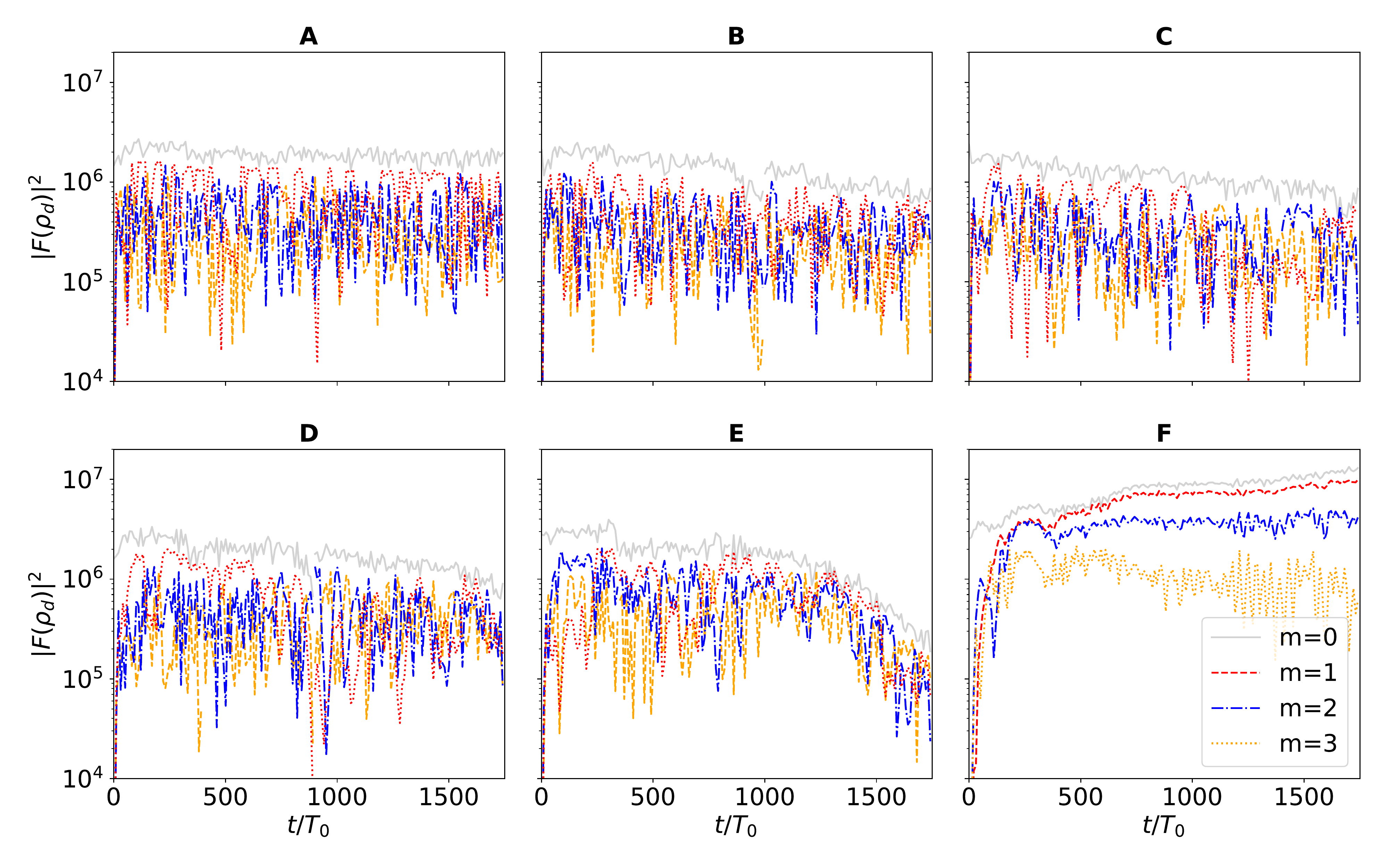}}

\medskip
\noindent {\bf Fig. S7. Power spectra for the PEI model of the debris disc at SDSS\,J1228+1040.} Time series in units, $T_0$ - where $T_0$ is the orbital period at the reference radius $r=1 R_{\odot}$ - of the power spectrum, $|F(\rho_d)|^2$ of the different arc systems shown in Fig.\,S6 A, broken down by wavenumber. The lettering a-f of each system in Fig.\,S6 corresponds to A-F here. Most of the power is in the $m=0$ mode, but substructure is apparent as power in the other modes. All systems (except f) show an alternation between $m=1$ and $m=2$ as dominant substructure, with $m=3$ at times in the cases of c, d and e. We conclude that no structure in these arcs systems is long lived enough to explain the variability observed in the Ca\,{\textsc{ii}} triplet. The system f results from boundary conditions and is still intensifying at the end of the simulation.

\clearpage
\centerline{\includegraphics[width=11cm]{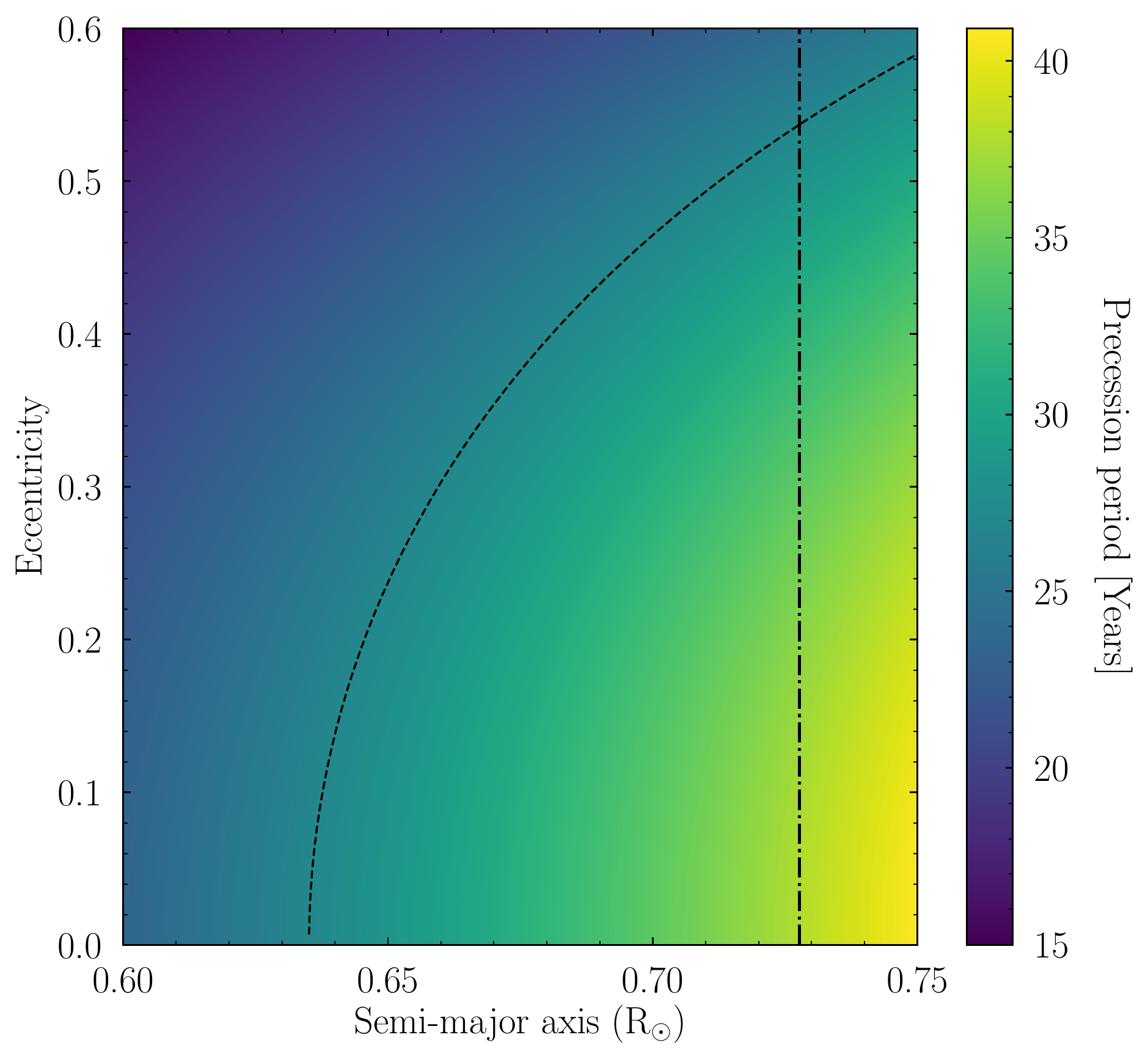}}

\medskip
\noindent {\bf Fig. S8. General relativistic precession periods.} The precession period of a body due to general relativistic effects in the gravitational field of the white dwarf in SDSS\,J1228+1040 is shown as a function of semi-major axis and eccentricity. The straight dot-dashed line indicates the semi-major axis 0.73\,R$_{\odot}$ for which a body will orbit on a period of 123.4\,min. The curved dashed line follows a precession period of 27\,yr, which is the precession period calculated from the long-term variability in the Ca\,{\textsc{ii}} triplet emission of the gaseous debris disc. These two curves cross at an eccentricity of $e$\,$\simeq$\,0.54.

\clearpage

\noindent{\bf Table S1. Log of observations.} The start and end times are given in units of Modified Julian Date (MJD).

\medskip

\begin{table}[h]\label{t-log}
\centering
\begin{tabular}{lccccc}
\hline
Date & Number of & Exposure & Start time & End time & Cycles\\
 & exposures & time [s] & [MJD] & [MJD] & Observed\\
\hline
2017--04--20 & 228 & 80 & 57863.85946 & 57864.12450 & 3.09\\
2017--04--21 & 131 & 120 & 57864.85940 & 57865.06820 & 2.44\\
2018--03--19 & 64 & 180 & 58196.08872 & 58196.24237 & 1.79\\
2018--04--10 & 32 & 180 & 58218.00653 & 58218.07959 & 0.85\\
2018--05--02 & 64 & 180 & 58240.90326 & 58241.05215 & 1.74\\
\hline
\end{tabular}
\end{table}

\clearpage

\noindent {\bf Table S2. The average equivalent width measurements of the Ca\,{\textsc{ii}} triplet in SDSS\,J1228+1040.} The measurements are given for each component of the Ca\,{\textsc{ii}} triplet for each night along with their uncertainties in brackets. Each profile is labeled by their zero-point air-wavelenth.

\medskip

\begin{table}[h]
\centering
\begin{tabular}{lrrr}
\hline
 & \multicolumn{3}{c}{Ca\,{\textsc{ii}} triplet profiles}\\
Date & 8498.020 [\AA] & 8542.090 [\AA] & 8662.140 [\AA] \\
\hline
2017--04--20 & -20.11 (0.02) & -28.31 (0.02) & -25.32 (0.02)\\
2017--04--21 & -19.70 (0.02) & -27.78 (0.02) & -24.93 (0.02)\\
2018--03--19 & -20.66 (0.02) & -28.66 (0.02) & -24.90 (0.03)\\
2018--04--10 & -18.26 (0.05) & -24.70 (0.04) & -21.40 (0.06)\\
2018--05--02 & -19.13 (0.02) & -26.33 (0.02) & -22.76 (0.02)\\
\hline
\end{tabular}
\end{table}

\clearpage

\noindent {\bf Table S3. Period determination of the short term variability.} Periods and uncertainties associated with the three strongest aliases in the amplitude spectrum computed from the 2017 equivalent width and blue-to-red ratio measurements (Fig.\,S3). The values in brackets are the likelihood of the individual aliases representing the true period of the short-period variability, assessed with a bootstrap test.

\medskip

\begin{table}[h]
\centering
\begin{tabular}{lrrr}
\hline
Data & \multicolumn{3}{c}{Period [min]}\\
\hline
2017 equivalent widths & 113.21$\pm$0.20~(0.2\%) & 122.88$\pm$0.19~(98.6\%) & 134.23$\pm$0.22~(1.2\%)\\
2017 blue-to-red ratio & 114.04$\pm$0.16~(0.0\%) & 123.63$\pm$0.15~(98.0\%) & 135.01$\pm$0.19~(2.0\%) \\
\hline
\end{tabular}
\end{table}

\end{document}